\documentclass[bg, hvmath]{copernicus}
\usepackage[usenames,dvipsnames]{xcolor}
\usepackage{booktabs}       
\usepackage{mathtools}      
\usepackage[redefsymbols=false,
            detect-weight = true,
            detect-inline-weight=math]{siunitx}        
\DeclareSIUnit{\year}{yr}
\DeclareSIUnit{\month}{mo}
\DeclareSIUnit{\molar}{M}

\definecolor{forestgreen}{rgb}{0.13, 0.55, 0.13}
\definecolor{amber}{rgb}{1.0, 0.75, 0.0}

\DeclareRobustCommand{\DutchName}[4]{#2~#1}

\begin{document}\hack{\sloppy}
\def\introductionexists{true}
\def\conclusionsexists{true}

\nolinenumbers      

\title{Manganese in the West Atlantic Ocean in context of the first global
ocean circulation model of manganese}
\Author[1,6]{Marco}{van~Hulten}
\Author[2,3,4]{Rob}{Middag}
\Author[1]{Jean-Claude}{Dutay}
\Author[4,5]{Hein}{de~Baar}
\Author[1]{Matthieu}{Roy-Barman}
\Author[1]{Marion}{Gehlen}
\Author[7]{Alessandro}{Tagliabue}
\Author[6]{Andreas}{Sterl}

\affil[1]{Laboratoire des Sciences du Climat et de l'Environnement (LSCE), IPSL,
CEA--Orme des Merisiers, 91191 Gif-sur-Yvette, France}
\affil[2]{Department of Chemistry, NIWA/University of Otago Research Centre for
Oceanography, Dunedin 9054, New Zealand}
\affil[3]{Department of Ocean Sciences \& Institute of Marine Sciences,
University of California Santa Cruz, CA 95064, USA}
\affil[4]{NIOZ Royal Netherlands Institute for Sea Research, Department of Ocean
Systems, and Utrecht University, P.O.~Box 59, 1790~AB Den Burg, Texel, the
Netherlands}
\affil[5]{University of Groningen (RUG),
Postbus 72, 9700~AB Groningen, the Netherlands}
\affil[6]{Royal Netherlands Meteorological Institute (KNMI),
Utrechtseweg 297, 3731~GA De Bilt, the Netherlands}
\affil[7]{University of Liverpool,
4 Brownlow Street, Liverpool L69~3GP, UK}

\runningtitle{Manganese in the Atlantic Ocean}
\runningauthor{M.M.P.~van~Hulten et~al.}
\correspondence{M.~M.~P.~van~Hulten (\texttt{mvhulten@lsce.ipsl.fr})}


\firstpage{1}
\maketitle


\begin{abstract}
Dissolved manganese (Mn) is a biologically essential element.
Moreover, its oxidised form is involved in removing itself and several other trace
elements from ocean waters.
Here we report the longest thus far 17\,500\;km length full-depth ocean section
of dissolved Mn in the West Atlantic Ocean, comprising 1320 data values of high
accuracy.
This is the GA02 transect that is part of the \textsc{Geotraces}
programme, which aims to understand trace element distributions.
The goal of this study is to combine these new observations with a new,
state-of-the-art, modelling to give a first assessment of the main sources and
redistribution of Mn throughout the ocean.
To this end, we simulate the distribution of dissolved Mn using a
global-scale circulation model.
This first model includes simple parameterisations to account
for the sources, processes and sinks of Mn in the ocean.
Oxidation and (photo)reduction, aggregation, settling, as well as biological
uptake and remineralisation by plankton, are included in the model.
Our model provides, together with the observations, the following insights:
\begin{itemize}
\item The high surface concentrations of manganese are caused by the
combination of photoreduction and sources to the upper ocean.
The most important sources are sediments, dust, and, more locally,
rivers.
\item Observations and model simulations suggest that surface Mn in the Atlantic
Ocean moves downwards into the southward flowing North Atlantic Deep Water
(NADW), but because of strong removal rates there is no elevated concentration
of Mn visible any more in the NADW south of 40\degree\,N.
\item The model predicts lower dissolved Mn in surface waters of the
Pacific Ocean than the observed concentrations.
The intense Oxygen Minimum Zone (OMZ) in subsurface waters is deemed to be a
major source of dissolved Mn also mixing upwards into surface waters, but the
OMZ is not well represented by the model.
Improved high resolution simulation of the OMZ may solve this problem.
\item There is a mainly homogeneous background concentration of dissolved Mn of
about \SIrange{0.10}{0.15}{\nano\molar} throughout most of the deep ocean.
The model reproduces this by means of a threshold on particulate manganese
oxides of \SI{25}{\pico\molar}, suggesting that a minimal concentration of
particulate Mn is needed before aggregation and removal become efficient.
\item The observed distinct hydrothermal signals are produced by assuming both a
strong source and a strong removal of Mn near hydrothermal vents.
\end{itemize}
\end{abstract}

\introduction                  \label{sec:mang:intro}

Dissolved manganese (\chem{Mn_\text{diss}}) is taken up by phytoplankton,
because Mn is crucial for photosynthesis and other biological functions
\citep{raven1990}.
Furthermore, its oxidised form
(\chem{Mn_\text{ox}}) plays an important role in the removal of several
other trace metals from seawater \citep{yamagata1963}.
While in the open ocean manganese (Mn) exists in small concentrations, it is the
twelfth most plentiful element in the Earth's crust \citep{wedepohl1995}.
In seawater, Mn occurs in many forms, among which the bioavailable dissolved
form.
After phytoplankton death, incorporated Mn sinks downwards
together with the dead material, but most of the organic material is
remineralised before reaching the sea floor \citep{froelich1979},
releasing Mn back to the water.

Another important mechanism of storing Mn in particles, besides biological
incorporation, is the removal of dissolved Mn via larger colloids on which the Mn
is oxidised to insoluble Mn(IV\@) (and possibly the other Mn(III) oxidation
state), and the subsequent aggregation by particulate matter.
Oxidation
occurs everywhere in the ocean where oxygen is available.
This process can be
strongly accelerated by Mn(II)-oxidising microorganisms, primarily bacteria and
fungi \citep{sunda1988,sunda1994,tebo2005}, but we do not understand the role of
these organisms quite well \citep{nealson2006}.
The reverse process is the reduction of Mn oxides to bioavailable
dissolved Mn(II), i.e.\ Mn$^{2+}\text{({\it aq})}$.
The full oxidation/reduction (redox) equilibrium reaction, in its most simple
form, is given by \citep{froelich1979}:
\begin{reaction}
    \chem{ Mn^{2+} + \frac{1}{2} O_2 + H_2O }
    \hspace{3mm}\leftrightharpoons\hspace{3mm}
    \chem{ MnO_2 + 2 H^+ } \,.
    \label{rcn:redox}
\end{reaction}
Reduction is significantly faster under the influence of sunlight.
It is hence referred to as \emph{photoreduction} when irradiance is the major
contributor.
The relative rate of reduction compared to oxidation
is important for Mn(II) availability.
An overall net higher oxidation rate implies more particle formation, hence more Mn export.

Manganese enters the open ocean through lithogenic dust deposition \citep{baker2016} and
lateral advection from reducing sediments \citep{homoky2016}.
Sediments along the relatively shallow (<1000\;m depth) ocean margins tend to
receive more organic deposition, hence by enhanced microbial decomposition are
more anoxic and a stronger source of reduced Mn than deep sea sediments.
Surface \chem{[Mn_\text{diss}]} is especially high in the central Atlantic Ocean and up
to at least 30\degree\,N
because of high dust input from the Sahara in combination with photoreduction
\citep{landing1987,jickells1995,guieu1994,baker2006,dejong2007,wu2014}.
Similarly,
\chem{[Mn_\text{diss}]} is high in the northern Indian Ocean \citep{vu2013}.
Dissolved Mn diffuses out of oxygen deprived sediments, because sediment microorganisms reduce
\chem{Mn_\text{ox}} if there is no more oxygen or nitrate left
\citep{li1969,landing1980,sundby1985,middag2012}.
As long as there is oxygen in the sediment, the organic carbon is remineralised by
using this oxygen as an electron acceptor \citep{froelich1979}.

Rivers are another source of Mn to the ocean \citep{elderfield1976,aguilar2006}.
While much of the fluvial Mn is removed within the estuaries into the sediments by scavenging and
aggregation, a large part may finally be transported to the ocean by diffusion
from and resuspension of the sediments \citep{jeandel2016,charette2016}.
Typically, the smaller sediment particles (\SIrange{0.5}{4}{\micro\metre})
have a high Mn content, and, because of their small size, are able to reach
the open ocean \citep{yeats1979,sundby1981,trefry1982}.
Manganese may also flux into the ocean by melting sea ice \citep{middag2011:mn:arctic}.
Finally, overwhelming evidence is found of manganese fluxing out of hydrothermal vents
\citep{klinkhammer1977,klinkhammer1985,hydes1986,klinkhammer2001,middag2011:mn:southern,middag2011:mn:arctic,german2016}.

Downward fluxes of settling particles that have been collected in sediment traps show a
strong correlation between lithogenic particles and authigenic Mn
\citep{roybarman2005}.
Therefore, lithogenic particles are likely to play a
significant role in the removal (oxidation, scavenging and aggregation) of Mn
\citep{dutay2015}.
Here the ballast effect of lithogenic particles, which typically have a density
of about twice of that of
seawater, is likely playing a major role for rapid settling of
agglomerates of biogenic and lithogenic particles including Mn-oxides coatings
\citep{balistrieri1981}.
This is consistent with the fact that most suspended particles are small (less
than \SI{2}{\micro\metre}), but the larger aggregates are deemed to be the
significant contributors to the vertical flux \citep{mccave1975,bishop1987}.
The complete process may be more complicated than described above
\citep{boyle2005}, e.g.\ because Mn binds to dissolved ligands such that more of it may
stay in solution \citep{sander2011,madison2013,luther2015}.
Manganese oxides are important scavengers of other trace metals like iron,
cobalt, nickel and zinc \citep{yamagata1963,murray1975,means1978,saito2004,tonkin2004}, as well as
insoluble radionuclides such as thorium and protactinium
\citep[e.g][]{reid1979,hayes2015:scavenging,jeandel2015}.
Therefore, Mn availability does not only directly impact primary production but
may also play a role in removing other elements from the surface ocean.
These elements include biologically essential trace metals as well as
many more trace elements.

Published observational studies, like the results of \citet{wu2014},
show low, relatively constant \chem{Mn_\text{diss}} concentrations of around
\SI{0.15}{\nano\molar} away from the boundaries.
Towards the surface there is a sharp increase of~\chem{[Mn_\text{diss}]}. 
Other notable elevations are near oceanic ridges.
These elevations of \chem{[Mn_\text{diss}]} at mid-depths can be ascribed to
hydrothermal activity near those regions.
The hydrothermal plumes typically extend in the order of \SI{1000}{\kilo\metre}
\citep{middag2011:mn:southern,middag2011:mn:arctic}, or reach even up to
\SI{3000}{\kilo\metre} \citep{resing2015}.

What makes the distribution of \chem{Mn_\text{diss}}{} relatively homogeneous in
the interior of the ocean?
How can this be reconciled with localised features where \chem{[Mn_\text{diss}]} is
very elevated compared to that stable ``background concentration''?
%
Until now only local model simulations of the Mn ocean cycle have been
performed that focus on the processes most relevant for the respective
regions.
In other words, the different features of the \chem{Mn_\text{diss}} distribution
have not been brought together in a unified model.

The case of hydrothermal activity has been studied by
\citet{lavelle1992} who modelled Mn in the deep ocean near hydrothermal
vents.
They included four Mn tracers, namely, a dissolved form, small particles
associated with bacteria, larger aggregate particles, and one in
sediments as the model includes benthic fluxes.
They found that ``more than 80\,\% of the hydrothermal Mn is deposited
within several hundred kilometres of the ridge crest though dissolved Mn
concentrations beyond that distance exceed background levels by many
times''.
This illustrates a high Mn input and removal from hydrothermal vents, as well
as the already mentioned large plume extend.

The \chem{Mn_\text{diss}} distribution in the North Pacific Ocean has
been modelled by \citet{johnson1996}, in a 1-D vertical model neglecting
horizontal transport.
Their oxidation model depends on the oxygen concentration and the hydroxide activity.
Their region of interest was on the upper and intermediate depth ocean,
and their goal was to reproduce the \chem{[Mn_\text{diss}]} maximum in
the Oxygen Minimum Zone (OMZ\@).
Besides the North Pacific, the OMZ is also present in other basins, amongst which 
the northwestern Indian Ocean \citep{saager1989,lewis2000}.
\citet{johnson1996} found that the combination of remineralisation rates and
decreased oxidation in the OMZ\@ explained the \chem{[Mn_\text{diss}]}
subsurface maximum in their profiles, whereas a flux from the continental margin
sediments appeared not responsible.
In the euphotic zone \chem{Mn_\text{diss}} was incorporated in
phytoplankton, while remineralised \chem{Mn_\text{diss}} in the aphotic
(and disphotic) zone was lost by oxidation and scavenging.

While these modelling efforts are useful for their purposes, no studies
exist in which the global ocean \chem{Mn_\text{diss}} distribution is modelled.
To arrive at a more integrated understanding of the behaviour of pelagic Mn,
we include a Mn model in a global ocean general circulation model.
Specifically, we will test a simple mechanism that should be able to give
insight into the apparently stable \chem{Mn_\text{diss}} concentration, and its
contrast to strong Mn sources.
This is the first time that a global ocean model for manganese has been written and
assessed.
It is a basic model that should give a starting point for further studies.
At the moment there is no mechanistic evidence for typical
uptake-remineralisation processes as is the case for, e.g., iron.
However, as there is of course uptake of Mn by phytoplankton, and sometimes it
can even be a limiting nutrient \citep{middag2013:weddell}, we include a biological
cycle of Mn in the model.
While we perform a model simulation on a global scale, we give more
attention to the \chem{Mn_\text{diss}} distribution in the Atlantic Ocean.
We will compare our simulations in most detail, and quantitatively, with a
here reported highly accurate dataset from the \textsc{Geotraces} programme,
namely the GA02 section in the West Atlantic Ocean.
Here \emph{accuracy} refers to the proximity of the measurements to the true
values, as confirmed by (i)~the compliance of reference samples with
international consensus values (Table~1), and (ii)~the agreement at the crossover
station (Fig.~\ref{fig:MnBermuda}) with the US section GA03, and (iii)~the
agreement of the two independent methods FIA and ICPMS for the our data of
section GA02 (Fig.~\ref{fig:correlationMn}).
In addition, the North Atlantic GA03 \citep{wu2014} and the Zero Meridian
Southern Ocean GIPY5 \citep{middag2011:mn:southern} transects will be used for
further detailed visual comparison.
Furthermore, in those regions we can study important properties of the ocean
geochemistry of manganese and the interaction with circulation, among which the
Atlantic overturning circulation and hydrothermal activity.

In this study our goal is to assess the fundamental processes that are the most
important to accurately simulate \chem{[Mn_\text{diss}]}, including
the aforementioned properties of the dissolved Mn distribution.
In this case, \emph{accuracy} refers to the proximity of the model to the
observations.
For clarity we will always qualify if this relates to the closeness of the model
simulation to the observations instead of the proximity of the measurements to
the true values.
To this end, we will first introduce our Mn model with its processes, sources and sinks.
We will show the results of a reference simulation, which will be
compared with recent high-accuracy observations.
Also several sensitivity simulations will be presented, studying the effects of
the biological cycle, the intense nature of hydrothermal vents and the strong
removal of manganese from the seawater.

\section{Methods}

\subsection{Observations in the West Atlantic Ocean}             \label{sec:data}

\subsubsection{Sample collection}          \label{sec:data:sampling}
For the determination of trace metal concentrations,
samples were collected along the \textsc{Geotraces} Atlantic
Meridional GA02 transect of the Netherlands (Fig.~\ref{fig:cruises_Mn}).
Sampling was done with an all-titanium ultraclean CTD sampling
system for trace metals \citep{debaar2008:titan} with novel PVDF samplers
\citep{rijkenberg2015,middag2015:Al,middag2015:BATS}.
Immediately upon recovery, the complete titanium frame with its 24 PVDF samplers
was placed inside a clean room environment where the sub-samples
for trace metal analysis were collected. The water was filtered from the PVDF
samplers over a \SI{0.2}{\micro\metre} filter cartridge (Sartobran-300,
Sartorius) under pressure (1.5\;atm) of (inline prefiltered) nitrogen gas.
Sub-samples for dissolved metals were taken in cleaned \citep[for cleaning
procedure]{middag2009} LDPE sample bottles. All sample bottles were rinsed five
times with the sample seawater. Seawater samples were acidified with HCl to a
concentration of 0.024\;M HCl which results in a pH of 1.7 to~1.8 with
Baseline\textregistered{} Hydrochloric Acid (Seastar Chemicals Inc.).

\subsubsection{Analysis of dissolved Mn}   \label{sec:data:analysis_Mn}
Analyses of dissolved manganese were performed shipboard with the method
developed by \citet{doi2004} with some slight modifications in the preparation
and brands of the chemicals used.
Notably, samples were buffered in-line
with an ammonium borate sample buffer to a pH of $8.5 \pm 0.2$.
This buffer was produced by dissolving
30.9\;g of boric acid (Suprapure, Merck) in 1\;L MQ water (Millipore Milli-Q)
deionised water $R > \SI{18.2}{\mega\ohm\per\centi\metre}$ and adjusting the pH
to 9.4 with ammonium hydroxide (Suprapure, Merck).

The buffered sample was pre-concentrated during 150\;s on a Toyopearl AF-Chelate
650M (TosoHaas, Germany) column. Hereafter the column was rinsed for 60\;s with
MQ water to remove interfering salts. The Mn was
subsequently eluted from the column for 200 seconds with a solution of 0.1\;M
three times quartz distilled formic acid (reagent grade, Merck) containing
0.1\;M hydrogen peroxide (Suprapure, Merck) and 12\;mM ammonium hydroxide
(Suprapure, Merck). The pH of this carrier solution was adjusted to $2.9 \pm
0.05$. The eluate with the dissolved Mn passed a second column of immobilised
8-hydroxyquinoline \citep{landing1986} to remove interfering iron ions in the
carrier solution \citep{doi2004}. Hereafter the carrier mixed 
with 0.7\;M ammonium hydroxide (Suprapure Merck) and a luminol solution. The
latter luminol solution was made by diluting \SI{600}{\micro\litre} luminol
stock solution and \SI{10}{\micro\litre} TETA (triethylenetetramine, Merck) in
\SI{1}{\cubic\deci\metre} MQ\@. The luminol stock solution was made by diluting 270\;mg luminol
(3-aminophtalhydrazide, Aldrich) and 500\;mg potassium carbonate in 15\;ml MQ\@.
The resulting mixture of carrier solution, ammonium hydroxide and luminol
solution had a pH of $10.2 \pm 0.05$ and entered a 3\;m length mixing coil
placed in a water bath of 25\;\degree{}C\@. Hereafter the chemiluminescence was
detected with a Hamamatsu HC135 Photon counter.  Concentrations of dissolved Mn
were calculated in nanomole per litre (nM) from the photon emission peak height
of triplicate measurements.

The system was calibrated using standard additions from a 5000\;nM Mn stock
solution (Fluka) to filtered acidified seawater of low Mn concentration that was
collected in the sampling region. A five-point calibration line (0, 0.1, 0.2, 0.6
and 1.2\;nM standard additions) and blank determination were made every day. The
three lowest points (0, 0.1 and 0.2\;nM) of the calibration line were measured
in triplicate and the two highest points (0.6 and 1.2\;nM) in duplicate in order
to add more weight to the lower part of the calibration line.  The blank was
determined by measuring acidified MQ which was below the detection limit and
subsequently no blank was substracted. The limit of detection defined as three
times the standard deviation of the lowest value observed was $< 0.01$\;nM. The
flow injection system was rinsed every day with a 0.5\;M HCl solution. 

An internal reference sample was measured in triplicate every day. This was a
sub-sample of a \SI{25}{\cubic\deci\metre} volume of filtered seawater that was taken at the
beginning of Leg~1 (also used during Leg~2, i.e.\ Iceland to the equator) and Leg~3
(Punta Arenas, Chile, to the equator). The relative standard deviation (i.e.\
the precision) of this replicate analysis seawater sample that was analysed 40
times on different days in triplicate was 2.57\,\% (Leg~1 and~2) and 1.21\,\%
for 17 analyses during Leg~3. The relative standard deviation on single days was
on average 1.37\,\% and the absolute values were 0.45 and 0.61\;nM for the first
two legs and Leg~3, respectively. 

\begin{table*}
\centering
\begin{tabular}{llll}
\toprule
\chem{Mn_\text{diss}}\ (\si{\nano\mole\per\kilo\gram}) & Consensus & \citet{middag2015:BATS} \\
\midrule
SAFe D2                             & $0.35 \pm 0.05  $   & $0.33\pm 0.01 \;(n=24)$ \\
\textsc{Geotraces} S                & $1.46 \pm 0.14  $   & $1.47\pm 0.03 \;(n=10)$ \\
\textsc{Geotraces} D                & $0.21 \pm 0.03  $   & $0.18\pm 0.01 \;(n= 5)$ \\
\bottomrule
\end{tabular}
\caption{Compliance with the international \textsc{Geotraces} reference samples program.
Left column are the international consensus values. Right column are the values
reported here as part of the GA02 dataset.}
\label{tab:consensus_Mn}
\end{table*}

As external comparison, the international reference samples collected on the
\textsc{Geotraces} Intercalibration Cruise (\url{www.geotraces.org}) as well as
from the SAFe cruise \citep{johnson2007} were analysed for Mn.
At different locations, large volumes of seawater were sampled of which
subsamples were analysed by various laboratories, resulting in different
reference values.
These labs include the Royal Netherlands Institute for Sea Research (NIOZ), who
also analysed the GA02 transect data.
An independent referee removed outliers and averaged the reference values,
resulting for each of the sample locations in a consensus value,
which is here considered as the true value.
These consensus values are listed in Table~\ref{tab:consensus_Mn}, together with
the values determined by NIOZ\@.
The distributions of the measurements from NIOZ lie within the precision of the
consensus values.

\begin{figure}[h] 
    \centering
    \includegraphics[width=.8\linewidth]{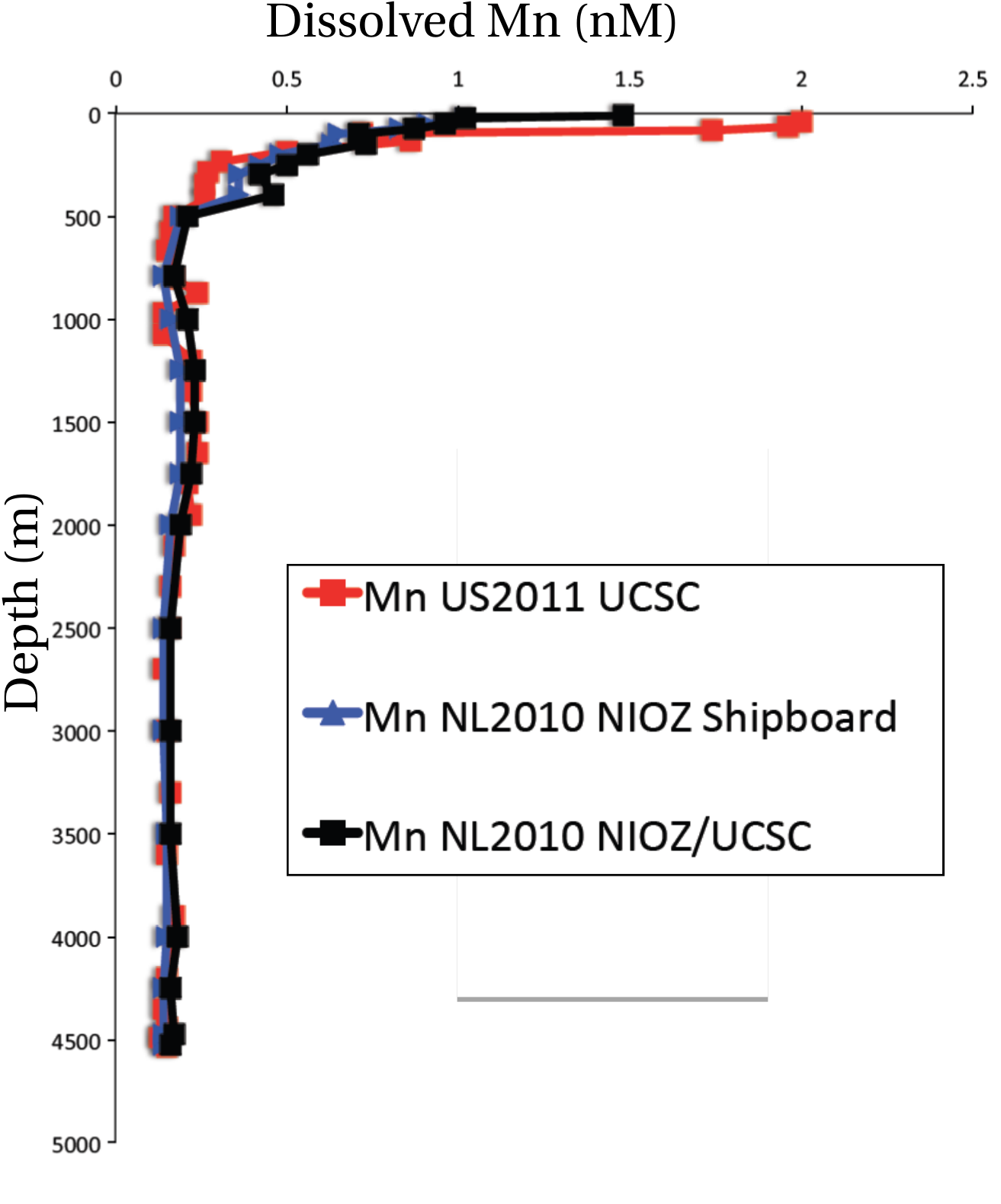}
    \caption{\chem{[Mn_\text{diss}]} at the 16~June 2010 (NL) and the
19~November 2011 (US) occupations of the crossover site of the Bermuda Atlantic
Time Series (BATS) station.
There is good agreement between the two sets of samples, as well as the two
different analytical methods, the shipboard FIA by NIOZ for the NL2010 samples,
and the lab ICP-MS at UCSC for both the NL2010 and the US2011 samples.
Each lab had its own independently prepared lab standards, further confirming
the overall accuracy.
Analyses by Rob Middag at NIOZ in 2010 and in 2011 at UCSC.}
    \label{fig:MnBermuda}
\end{figure}

The vertical profiles of the BATS station for the Netherlands sampling at
13~June 2010 and the US sampling at 19~November 2011 are presented in
Fig.~\ref{fig:MnBermuda}.
For both occupations the samples were analysed by ICP-MS by Rob Middag at UCSC\@.
Moreover, the same analyst had done shipboard FIA analyses during the 2010 cruise.
There was no statistical difference within analytical uncertainty between the
three datasets in the deep waters \citep{middag2015:BATS}.
Small differences within the upper $\sim$500\;m between the Netherlands sampling
and the US sampling are deemed to be real due to the seasonal variability over
the more than one year time difference of sampling.
Furthermore, these Mn profiles are also consistent with the profiles determined
by \citet{landing1995}.

\begin{figure}
    \centering
    \includegraphics[width=\linewidth]{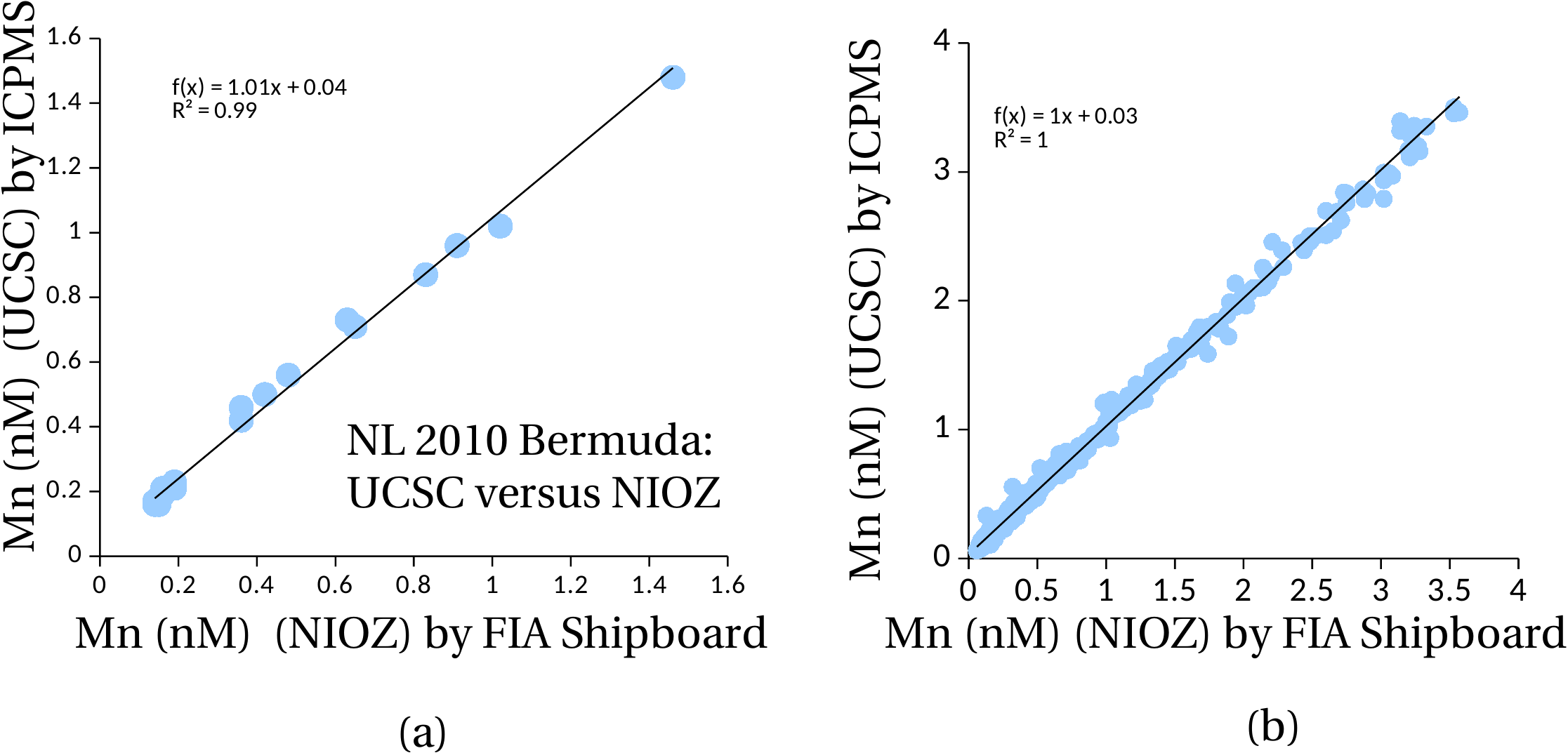}
    \caption{The correlation between the two methods of analysis for the
            determination of \chem{[Mn_\text{diss}]} at GA02 (shipboard and laboratory
            mass spectrometry): (a) at Bermuda; (b) all 55 West Atlantic stations.}
    \label{fig:correlationMn}
\end{figure}

Besides the shipboard Flow Injection Analysis (FIA), we also analysed samples
using mass spectrometer analysis.
Figure~\ref{fig:correlationMn} shows the correlation between the two methods of
analysis for the determination of \chem{[Mn_\text{diss}]}, shipboard and
laboratory measurements.
There is a very good agreement between the shipboard and mass spectrometer
analyses, which strongly suggests a high observational accuracy.
For the GIPY5 transect in the Southern Ocean, we analysed
\chem{[Mn_\text{diss}]} only through flow injection (FIA\@).
Since we want to plot GA02 together with the GIPY5 in this study, we use the
observational data obtained from the shipboard FIA for the comparison with the
model simulations.

\subsection{Model description}

In order to simulate the three-dimensional (3-D) distribution of dissolved Mn,
we use the general circulation model \textit{Oc\'ean PArall\'elis\'e} (OPA)
that is part of NEMO, a framework for ocean models \citep{madec2016}.
We use the ORCA2-LIM configuration of NEMO\@.
The spatial resolution is 2$^{\circ}$ by 2$^{\circ}\,\cos(\phi)$ (where $\phi$
is the latitude) with an increased meridional resolution to 0.5$^{\circ}$ in the
equatorial domain \citep{madec1996}.
The model has 30 vertical layers, with an increased vertical thickness from
\SI{10}{\metre} at the surface to \SI{500}{\metre} at \SI{5000}{\metre} depth.
Representation of the topography is based on the partial-step thickness
\citep{barnier2006}.
Lateral mixing along isopycnal surfaces is performed both on tracers and
momentum \citep{lengaigne2003}.
The parameterisation of \citet{gent1990} is applied from 10\degree{} poleward to
represent the effects of non-resolved mesoscale eddies.
Vertical mixing is modelled using the turbulent kinetic energy (TKE) scheme of
\citet{gaspar1990}, as modified by \citet{madec2016}.
The fluid dynamics used to drive our model is identical to that used in
\citet{aumont2015}.

The model contains two tracers of Mn, referred to as dissolved
(\chem{Mn_\text{diss}}) and oxidised (\chem{Mn_\text{ox}}) manganese.
These tracers are driven by the equations set out in this section, as well as 
the velocity fields obtained from OPA\@.
Instead of calculating the dynamical variables of the model (velocity and
mixing), we run it off-line, using a climatology with a resolution of five days.

\begin{figure}
    \centering
    \includegraphics[width=\linewidth]{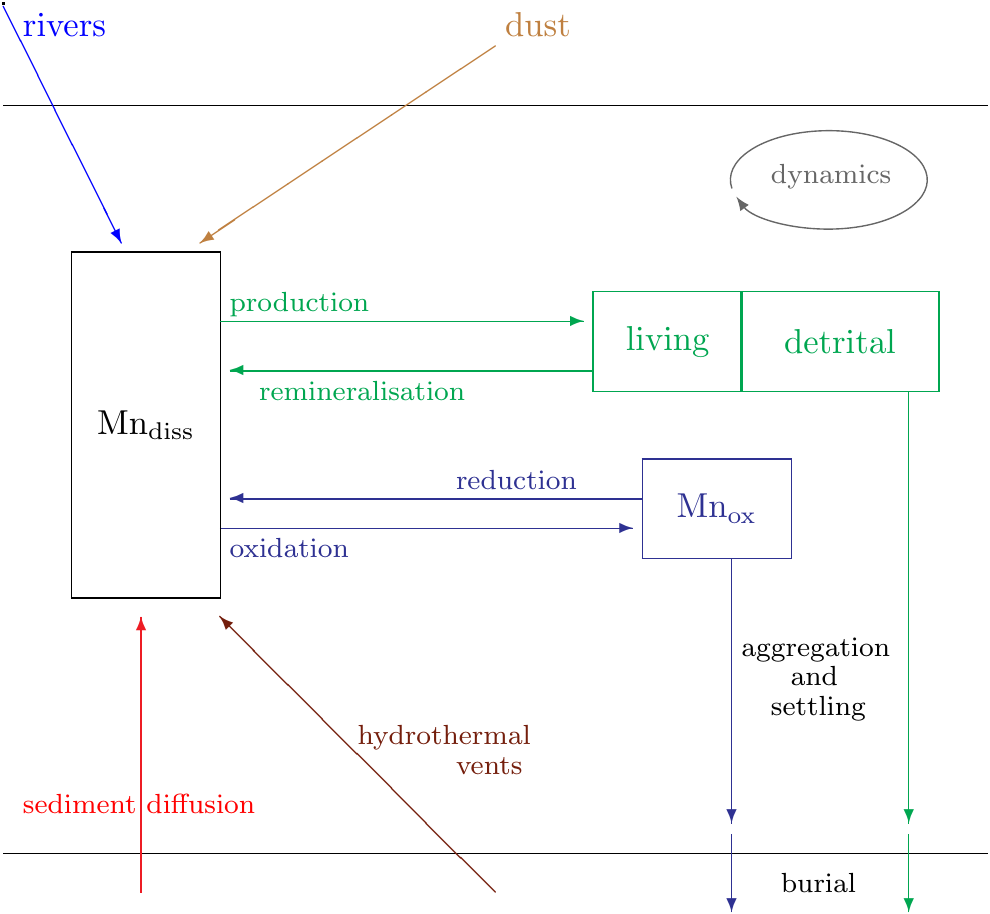}    
    \caption{Model scheme:
    biology in green; redox and scavenging in blue; circulation and mixing
    (dynamics) in grey; and the sources dust, rivers, hydrothermal and sediment
    are in light brown, light blue, dark brown and red, respectively.
    \chem{Mn_\text{diss}} is the dissolved and \chem{Mn_\text{ox}} the oxidised Mn.}
    \label{fig:mang:model_scheme}
\end{figure}

\begin{figure*}
    \centering
    \includegraphics[width=\linewidth]{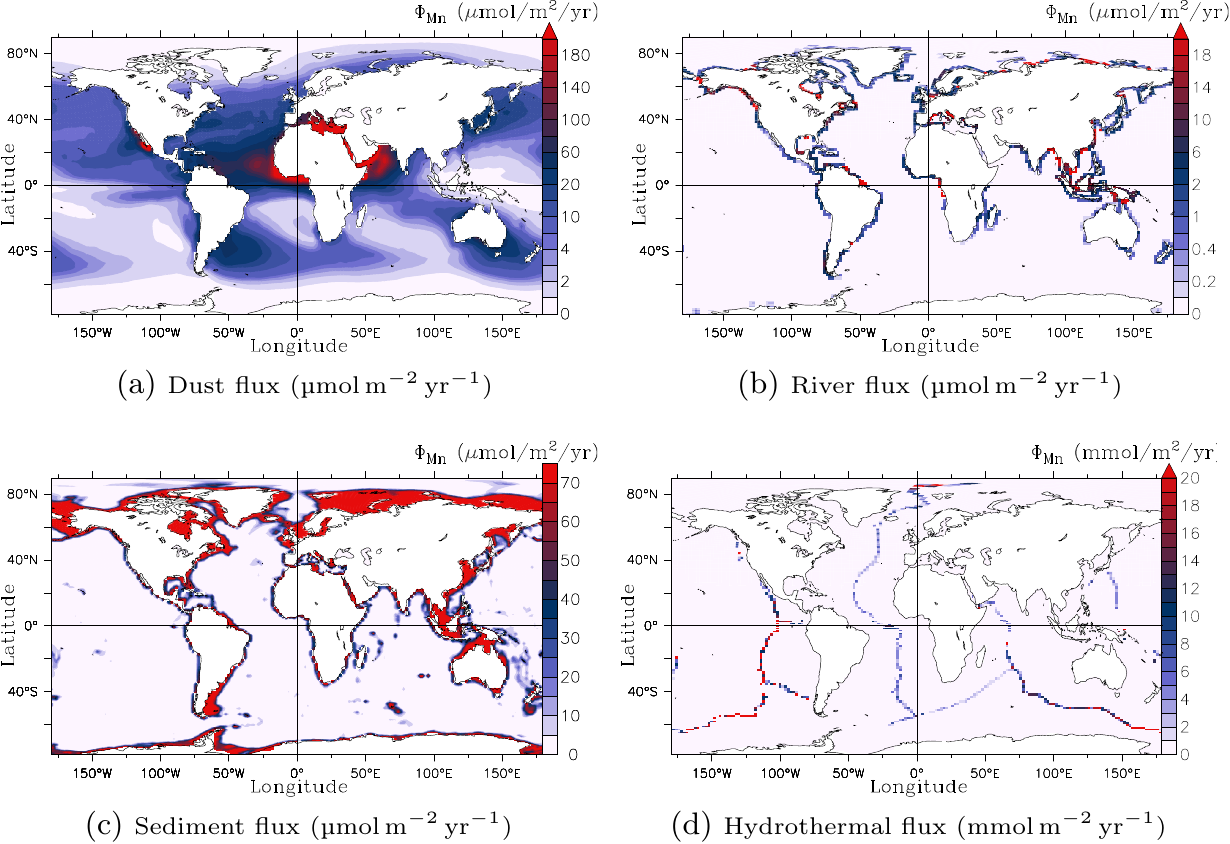}
    \caption{Sources of Mn to the ocean: effective \chem{Mn_\text{diss}} input flux~$\Phi$.
        Three-dimensional fields are vertically integrated, such that dimensions
        are molar fluxes.
        Ranges vary between the different sources.}
    \label{fig:mang:sources}
\end{figure*}

Figure~\ref{fig:mang:model_scheme} presents the conceptual scheme of our
manganese model.
The internal processes include biological uptake and remineralisation (green
arrows), reduction, oxidation, aggregation and burial (blue arrows in the
figure), and are described in the following subsections.
Manganese sources from rivers, the atmosphere, sediments and hydrothermal vents
are presented as arrows at the top and bottom of the figure.
These four Mn sources are presented in Fig.~\ref{fig:mang:sources}, and
Table~\ref{tab:mang:sources} lists the absolute
contributions to the different basins by each of these sources, as well as
the relative contribution of every source to the world ocean.
The model parameters are summarised in Table~\ref{tab:params_Mn}.
In the following subsections we will describe how the different sources and
processes are included in the model.

\begin{table*}
  \centering
  \begin{tabular}{lrrrr}
    \toprule
    Basin           & {Dust}    & {Rivers}  & {Sediment}& {Hydrothermal} \\ 
    \midrule
    Atlantic Ocean  & 2200      & 127       &  924      & 13917 \\ 
    Pacific Ocean   & 1183      &  94       & 1237      & 59846 \\ 
    Indian Ocean    & 1506      &  25       &  442      & 18653 \\ 
    Southern Ocean  &   14      &   0       &  206      &  7601 \\ 
    Arctic Ocean    &   23      &  19       &  463      &  2269 \\ 
    Mediterranean Sea & 673     &  12       &   91      &     0 \\ 
    \addlinespace
    total amount (Mmol\,yr$^{-1}$)
                    & 5598      & 277       & 3363      & 102286 \\ 
    relative amount (\%) & 5.0  & 0.2       & 3.0       & 91.7  \\ 
    \bottomrule
    \addlinespace
  \end{tabular}
  \caption{Absolute amount of effective annual input by an imposed Mn flux from
  sediments, rivers and hydrothermal vents into each basin
  (\si{\mega\mole\per\year}).
  The Southern Ocean is defined as the ocean south of 58.7\degree\,S.
  The line ``total amount'' denotes how much world-wide Mn is added to the
  ocean due to a specific flux; ``relative amount'' is normalised to the
  total Mn input flux.}
  \label{tab:mang:sources}
\end{table*}

\begin{table*}[t]
  \begin{tabular}{lrlll}
    \toprule
    Parameter                   & Symbol                    & Value used                & Known range       & Reference                             \\
    \midrule
    Mass fraction of Mn in dust & $f_\text{Mn,dust}$        & \num{880}\;ppm            & \numrange{696}{880}\;ppm & \citet{wedepohl1995};\\&&&&\citet{mendez2010}    \\
    Dust Mn solubility          & $\alpha$                  &  \num{40}\,\%             & \numrange{10}{70}\,\%    & \citet{baker2006}                  \\
    Sediment source Mn/Fe ratio & $r_\text{Mn:Fe,sed}$      &   \num{0.2}               & uncertain             & \citet{bortleson1974};\\&&&&\citet{slomp1997}    \\
    River source Mn/Fe ratio    & $r_\text{Mn:Fe,riv}$      &   \num{0.214}             & uncertain             & \citet{book::sarmiento2006}\\
    Hydrothermal Mn/\chem{^3\!He} ratio & $r_{\chem{Mn:^3\!He},\text{hydro}}$ & \num{0.10e9} & uncertain        & -                                  \\
    \addlinespace
    Settling speed of \chem{Mn_\text{ox}} & $w_\text{ox}$   & \num{1}--\SI{10}{\metre\per\day} & \num{0.9}--\SI{1.4}{\metre\per\day} & \citet{roybarman2009}\\
    Oxidation rate constant     & $k_\text{ox}$             & \SI{0.341e-3}{\per\hour}  & uncertain                           & \citet{bruland1994};\\&&&&\citet{sunda1994} \\
    Photoreduction rate constant& $k_\text{red,light}$      & \SI{98e-3}{\per\hour}     & \num{50}--\SI{150e-3}{\per\hour}    & \citet{sunda1994}    \\
    Aphotic reduction rate      & $k_\text{red,dark}$       & \SI{1.70e-3}{\per\hour}   & \num{0.98}--\SI{14.3e-3}{\per\hour} & \citet{bruland1994};\\&&&&\citet{sunda1994} \\
    Aggregation threshold       & $\mathcal{X}_\text{thr}$  & \SI{25}{\pico\molar}      & hypothetical                        & -                    \\
    Mn/P incorporation ratio    & $R_\text{Mn:P}$           & \num{0.36e-3}             & \numrange{0.2e-3}{1.5e-3}          & \citet{middag2011:mn:southern};\\&&&&\citet{twining2013}  \\
    \bottomrule
    \addlinespace
  \end{tabular}
  \caption{Mn model parameters for the \textit{Reference} simulation.
  The settling velocity $w_\text{ox}$ is given by Eq.~\ref{eqn:settlingMnox_step}.
  In \textit{LowHydro} the hydrothermal flux and the maximum settling speed are
  both reduced by a factor of 10 ($r_{\chem{Mn:^3\!He}\text{,hydro}} =
  \num{0.01e9}$; and $w_\text{ox} \equiv \SI{1}{\metre\per\day}$ or,
  equivalently, $\mathcal{X}_\text{thr} \rightarrow \infty$).
  In the sensitivity simulation \textit{NoThreshold} the aggregation threshold
  $\mathcal{X}_\text{thr}$ is set to zero.}
  \label{tab:params_Mn}
\end{table*}

\subsubsection{Atmospheric source}
Manganese is added to the pool of \chem{Mn_\text{diss}} in the upper model layer, according~to
\begin{equation}
    \frac{\partial \mathcal{D}}{\partial t}\Big|_\text{dust}
      = \frac{\alpha \cdot f_\text{Mn,dust}}{m \cdot \Delta z_1}
        \cdot \Phi_\text{dust} \,,
\end{equation}
where $\mathcal{D}$ is the dissolved Mn concentration, $\alpha$ is
the solubility of Mn in dust, $f_\text{Mn,dust}$ is the mass fraction of Mn
in dust, $m$ is the molar mass of Mn, and $\Delta z_1 =
\SI{10}{\metre}$ is the upper model layer thickness.
The lithogenic dust deposition flux, $\Phi_\text{dust}$, is derived from
the Interaction with Chemistry and Aerosols (INCA) model \citep{hauglustaine2004}.
Here we use a 12 month climatology of INCA's output as a forcing.

The average mass fraction of Mn in the Earth's upper
crust is 527\;ppm \citep{wedepohl1995}.
However, the fraction measured in Saharan dust is 880\;ppm
\citep{mendez2010}, consistent with \citet{guieu1994} and
\citet{statham1998}.
Since most of the dust deposited on the Atlantic Ocean originates from
the Sahara and our focus is the Atlantic Ocean, the value of 880\;ppm is
used for~$f_\text{Mn,dust}$.

The solubility of Mn from dust is uncertain and relatively high compared
to most other trace metals.
Here $\alpha = 40$\,\% of the Mn in dust is assumed to 
dissolve, largely consistent with the values reported by
\citet{guieu1994,jickells1995,baker2006,dejong2007,buck2010:solubility}.
Several studies report even higher values ($> 50$\,\%), which are,
however, mainly from anthropogenic or otherwise processed dust, while
the lower reported values ($< 50$\,\%) are from natural dust, mainly of
Saharan origin.

Figure~\ref{fig:mang:sources}a presents the average Mn dissolution flux
of the 12~months climatology.
Globally this is \SI{5.6}{\giga\mole\per\year} of~Mn.

\subsubsection{River source}
The manganese river source is modelled analogously to iron, which is part of the
biogeochemical model \textsc{Pisces} \citep{aumont2015}.
This means that our manganese influx is proportional to the total dissolved
(organic and inorganic) carbon flux, just like for iron \citep{aumont2015}.
Hence, the modelled concentration change of \chem{Mn_\text{diss}} caused by river input is
given by
\begin{equation}
    \frac{\partial\mathcal{D}}{\partial t}\Big|_\text{rivers} = 
        r_\text{Mn:Fe,river} \cdot \frac{\partial[\chem{Fe}_\text{diss}]}%
                                        {\partial t}\Big|_\text{rivers}\,,
    \label{eqn:mang:rivers}
\end{equation}
where $[\chem{Fe}_\text{diss}]$ is the dissolved iron concentration, and
$r_\text{Mn:Fe,river}$ is the effective manganese/iron
flux ratio, here set to 0.214, based on river dissolved concentrations
\citep[][p.~2]{book::sarmiento2006}.
For comparison, the Mn/Fe flux ratio is much higher than the crustal ratio,
which is around 0.02 \citep{wedepohl1995}, but probably underestimated as we do
not consider external sources of particulate manganese.
The effective \chem{Mn_\text{diss}} input into the ocean by rivers is
presented in Fig.~\ref{fig:mang:sources}b.
The global, effective river flux of Mn is \SI{0.28}{\giga\mole\per\year}.

\subsubsection{Sediment source}
The largest contribution of Mn to the upper ocean is dust deposition, but over
large shelf and slope regions (e.g.\ polar oceans) the flux of \chem{Mn_\text{diss}} from the
sediment can be of the same order of magnitude as, or higher than, the dust
deposition flux \citep[e.g.][for the Southern Ocean]{middag2013:weddell}.
The (redox) reactions in the sediment are not explicitly modelled, since
the sediment is not part of our model domain.
Therefore, Mn addition from the sediment is modelled as a prescribed
source.
Since we do not have global maps of \chem{Mn_\text{diss}} sediment--seawater flux, we
parameterise the flux based on existing parameterisations of nitrate and iron.

\citet{middelburg1996} derived an empirical model for calculating the
denitrification rate as a function of the seafloor depth:
\begin{equation}
    \zeta_\text{Fsed} = - 0.9543 + 0.7662 \cdot \ln(z_\text{Fsed})
                         - 0.235 \cdot \ln(z_\text{Fsed})^2 \,,
\end{equation}
where $\zeta_\text{Fsed}$ is the natural logarithm of the denitrification flux
(\si{\micro\mole\per\centi\metre\squared\per\day} of carbon) and $z_\text{Fsed}$
is a function of bathymetry.
\citet{aumont2015} used this model for their sediment source of dissolved Fe in
the \textsc{Pisces} model.
They used a high-resolution bathymetric map to account for the shallow shelves,
and modulated the seafloor depth ($z \rightarrow z_\text{Fsed}$) as follows:
\begin{equation}
    z_\text{Fsed} = \min\big(8, \big(\frac{z}{\SI{500}{\metre}}\big)^{-1.5}\big) \,.\\
\end{equation}
We set the maximum iron flux to \SI{1}{\micro\mol\per\metre\squared\per\day},
following \citet{aumont2006}. 
We assume a porewater ratio $r_\text{Mn:Fe,sed}=0.2$, and follow the same
method as \citet{aumont2015}.
Our final sediment addition of \chem{Mn_\text{diss}} into bottom water is given
by:
\begin{equation}
    \frac{\partial \mathcal{D}}{\partial t}\Big|_\text{sediments} = 
        \frac{\SI{0.2}{\micro\mol\per\square\metre\per\day}}{\Delta z_\text{sed}}
        \cdot \min(1,\; 2 \text{e}^{\zeta_\text{Fsed}})\,,
    \label{eqn:mang:sediment}
\end{equation}
where $z_\text{sed}$ is the gridbox thickness of the bottom gridbox (just above
the seafloor) in metre.
With this prescribed source, the Mn flux is limited to a maximum of about
\SI{75}{\micro\mole\per\metre\squared\per\year}.
This can be seen in Fig.~\ref{fig:mang:sources}c: the
higher-than-\SI{70}{\micro\mole\per\square\metre\per\year} regions on the
shelves is especially notable as it falls in the upper (red) part of
the colour scale.
The global sediment flux of Mn is \SI{3.4}{\giga\mole\per\year}.

\subsubsection{Hydrothermal source}
Hydrothermal vent \chem{Mn_\text{diss}} fluxes are modelled proportional to that of \chem{^3\!He}
in \citet{ruth2000} and \citet{dutay2004}.
This approach is shown to have worked for iron
\citep{bowers1988,douville2002,tagliabue2010,resing2015}.
The basic equation for the change of \chem{Mn_\text{diss}} from hydrothermal vent influx is
\begin{equation}
    \frac{\partial\mathcal{D}}{\partial t}\Big|_\text{hydrothermal} = 
        r_{\chem{Mn:^3\!He},\text{hydro}} \cdot \frac{\partial [\chem{^3\!He}]}%
                                        {\partial t}\,,
    \label{eqn:mang:hydro}
\end{equation}
where $r_\text{\chem{Mn:^3\!He},hydro} = \num{0.1e9}$ is the ratio
between the \chem{Mn_\text{diss}} and \chem{^3\!He} effective inflow
from hydrothermal vents into the model domain.
Recent observational studies found \chem{Mn_\text{diss}}/\chem{^3\!He}
concentration ratios of \num{4.0e6} \citep{kawagucci2008} and \num{3.5e6}
\citep{resing2015} at observational sites close to hydrothermal outflux regions.
To satisfy this, we assume that a dissolved fraction of 4\,\% is left when the
hydrothermal plume reaches these two observational sites.

This high ``solubility fraction'' means that the hydrothermal vents are a large source of
\chem{Mn_\text{diss}} compared to \chem{Fe} \citep{tagliabue2010,resing2015}.
The integrated Mn flux is \SI{102}{\giga\mole\per\year}.
In our simulations we will show that we need to assume such a large flux to
explain the observations.
This choice also relates to the fast modelled removal rate of \chem{Mn} near the
depth of hydrothermal vents (explained later).
The hydrothermal \chem{Mn_\text{diss}} source is presented in
Fig.~\ref{fig:mang:sources}d.

\subsubsection{Redox processes}
Reduction and oxidation of \chem{Mn} within the water column is a combination of several
processes.
Here we assume that \chem{Mn_\text{ox}} is subject to reduction with a constant
rate, but significantly stimulated by sunlight \citep[e.g.][]{sunda1994}.
This is taken into account in the model by using different $k_\text{red}$ for the
euphotic and aphotic zones of the ocean.
However, there are other processes playing a role that can locally be important.
Those include the microbial enhancement of the rate of oxidation in regions
where \chem{Mn_\text{diss}} supply is high \citep{sunda1994,tebo2005}, and the
dependence on the \chem{O_2} concentration and pH\@ \citep[for Fe at
GA02]{johnson1996,rijkenberg2014}.
However, at this stage we decide to not include a dependency on \chem{[O_2]} to
the model.
Hence, here we choose to model Mn following a pseudo-first-order reaction
where $k_\text{ox}$ and $k_\text{red}$ are pseudo-first-order rate
constants \citep[conventional primes omitted, e.g.][]{kinetics:stone1990}:
\begin{align}
    \frac{\partial\mathcal{D}}{\partial t}\Big|_{\text{redox}} &=
        -k_\text{ox} \mathcal{D} + k_\text{red} \mathcal{X} \\
    \frac{\partial\mathcal{X}}{\partial t}\Big|_{\text{redox}} &=
        k_\text{ox} \mathcal{D} - k_\text{red} \mathcal{X}  \,,
\label{eqn:redox}
\end{align}
where $\mathcal{X}$ is the particulate oxidised Mn concentration, and
\begin{equation}
    k_\text{red} = \begin{cases}
        k_\text{red,light} \;\;\;\; \text{in the euphotic zone} \\
        k_\text{red,dark} \;\;\;\; \text{elsewhere}    \,,
    \end{cases}
\end{equation}
where $k_\text{red,light}$ is the reduction rate in the euphotic zone and
$k_\text{red,dark}$ in the aphotic zone.
The euphotic zone is defined as the depths where the sunlight intensity is at
least~1\,\%.
The value for $k_\text{red,light}$ is taken from \citet{sunda1994} who found a
mean dissolution rate of natural Mn oxides of
$(98\pm23)\times\SI{e-3}{\per\hour}$.
The $k_\text{red,dark}$ is much smaller and much more uncertain (varying a
factor of fifteen in \citet{sunda1994}).
As we have both dissolved and particulate \chem{Mn} in our model, we chose to
fit the $k_\text{ox}/k_\text{red,light}$ to
\chem{[Mn_\text{ox}]/[Mn_\text{diss}]} from observations, resulting in a value
within the observational range of \citet{sunda1994}.
At the time of this study only \citet{bruland1994} was known to us as reporting
accurate dissolved and particulate Mn for the same samples (at the VERTEX-IV
station).
From this $k_\text{ox}$ was calculated, and as \citet{bruland1994} sampled the
deep ocean as well, $k_\text{red,dark}$ was derived from that study as well
(Table~\ref{tab:params_Mn}).


For one simulation, we will introduce a threshold on the oxidation process of
\chem{Mn_\text{diss}}.
In that case, oxidation only takes place when \chem{[Mn_\text{diss}]}.higher than a certain threshold
value $\mathcal{D}_\text{thr}$; in other words, $k_\text{ox}$ is multiplied by
$H(\mathcal{D} - \mathcal{D}_\text{thr})$.
The Heaviside step function $H(x)$ equals zero where $x\leq 0$ and one where $x>0$.
Based on observations, we have estimated $\mathcal{D}_\text{thr} =
\SI{0.125}{\nano\molar}$, corresponding to the observed deep ocean background
value away from the influence of hydrothermal sources.
The value can be reproduced by first sorting the \chem{Mn_\text{diss}}
concentrations.
Then we cut off the third quartile to remove the high values (above
\SI{0.42}{\nano\molar}), and chose a low value for a typical background
concentration (\SI{0.13}{\nano\molar} is the first quartile).
The reason for not choosing simply the minimum value is that values close to that
can be lower than the `background value' because of local removal processes.

\subsubsection{Settling and burial}
Manganese oxides settle, resulting in a concentration change according to
\begin{equation}
    \frac{\partial \mathcal{X}}{\partial t} \Big|_\text{settling}
        = - w_\text{ox} \cdot \frac{\partial\mathcal{X}}{\partial z} \,,
    \label{eqn:settlingMnox}
\end{equation}
where $w_\text{ox}$ is the settling velocity, set to a constant
\SI{1}{\metre\per\day} as long as $\chem{[Mn_\text{ox}]} \leq
\SI{25}{\pico\molar}$.
If $\chem{[Mn_\text{ox}]} > \SI{25}{\pico\molar}$, the settling velocity
is not a constant any more but a function of depth.
Still, in the mixed layer $w_\text{ox} = \SI{1}{\metre\per\day}$, but
if $\chem{[Mn_\text{ox}]} > \SI{25}{\pico\molar}$ the sinking speed
increases linearly such that it reaches a value of \SI{10}{\metre\per\day} at
\SI{2.5}{\kilo\metre} depth.
The manganese oxide is buried when arriving at the ocean floor, which
means that it is removed from the model domain.

The rationale for increasing the settling velocity
above a certain threshold is that if \chem{Mn_\text{ox}} is large enough, dense
aggregates form that have faster sinking rates.
This is not unlike increasing velocities of detritus in some models
\citep{kriest2011,aumont2015}.
Notably mineral particles (sand, clay, carbonate) of high density in the
order of 2--3 times that of seawater are responsible for this \citep{mccave1975}.
Particulate organic carbon may play an important role as well \citep{passow2006}.
In this way, \chem{[Mn_\text{diss}]} does not go to zero while still
providing a deep ocean sink.
As long as this critical particulate Mn concentration of \SI{25}{\pico\molar} is
not reached, aggregation of small \chem{Mn_\text{ox}} particles does not yet
occur.
The choice of the critical concentration of \SI{25}{\pico\molar}, also referred
to as the \emph{aggregation threshold}, is derived from the redox rate constants
in combination with a typical, low value of the observed dissolved Mn
concentration (here chosen as \SI{0.125}{\nano\molar}):
$\mathcal{X}_\text{thr}=(k_\text{ox}/k_\text{red,dark})\cdot\SI{0.125}{\nano\molar}=\SI{25}{\pico\molar}$.
More precisely, settling of \chem{Mn_\text{ox}} follows Eq.~\eqref{eqn:settlingMnox}
with the settling velocity
\begin{equation}
    w_\text{ox}/(\si{\metre\per\day}) = 1 + 9 \cdot
                \text{max}\left( 0, \frac{z-\mathit{MLD}}{\SI{2.5}{\kilo\metre}} \right)
                \cdot H(\mathcal{X} - \mathcal{X}_\text{thr}) \,,
    \label{eqn:settlingMnox_step}
\end{equation}
where $z$ is the depth, \textit{MLD} is the mixed layer depth, and
$\mathcal{X}_\text{thr}$ is the aggregation threshold set
to~\SI{25}{\pico\molar}.

\subsubsection{Biological cycle}
Manganese is incorporated into phytoplankton during growth.
To this end, we run the \textsc{Pisces}-v2 biogeochemical ocean model
\citep{aumont2015} together with our manganese model.
The biological processes where Mn is involved are modelled in proportion to the
change in phosphate concentration, thus it is given by:
\begin{equation}
\frac{\partial\mathcal{D}}{\partial t} \Big|_\text{biology} = R_\text{Mn:P}\cdot
\frac{\partial\chem{[PO_4^{3-}]}}{\partial t} \Big|_\text{biology} \,,
\label{eqn:bio}
\end{equation}
where \chem{[PO_4^{3-}]} is the phosphate concentration.
The extended Redfield ratio for Mn, $R_\text{Mn:P}$,
is set to \num{0.36e-3}, the value that was determined from data at
the GIPY5 Zero Meridian section \citep{middag2011:mn:southern}.
This is a typical value for the manganese to phosphorus ratio in phytoplankton,
though the full range of synchrotron X-ray fluorescence determined ratios,
i.e.\ those that correspond to intracellular concentrations, is
\numrange{0.2e-3}{1.5e-3} \citep{twining2013}.

There is no growth limitation of phytoplankton by shortage of dissolved~Mn,
i.e., in the model, manganese does not affect the biological carbon cycle.


There are four types of plankton in \textsc{Pisces}: nanophytoplankton,
diatoms, microzooplankton and mesozooplankton
(presented in Fig.~\ref{fig:mang:model_scheme} as the box ``living'').
There are two detrital pools, namely small particles settling with
\SI{2}{\metre\per\day} and large particles
settling in our model configuration with \SI{50}{\metre\per\day} (box
``detrital'').
Manganese is incorporated in the four living pools according to the
\textsc{Pisces} equations for phosphorus.
Through other processes they become part of detrital material or particulate
organic matter.
These pools are, however, not explicitly followed.
Only the biological sources and sinks of \chem{Mn_\text{diss}} are modelled,
entailing the conversion of \chem{Mn_\text{diss}} to the two phytoplankton
pools, and from all particle pools to \chem{Mn_\text{diss}}.


\subsection{Simulations}                \label{sec:mang:simulations}

The \textit{Reference} simulation was spun up for \SI{600}{yr} to reach a steady state.
From year~100 onwards two sensitivity simulations were forked off to run in
parallel with \textit{Reference} for another \SI{500}{\year}.
These simulations are variations of the \textit{Reference} simulation that uses
the parameters listed in Table~\ref{tab:params_Mn}.
Table~\ref{tab:mang:simulations} lists the simulations and their key parameters.

\begin{table*}[h]
  \begin{tabular}{l>{$}c<{$}>{$}c<{$}>{$}c<{$}>{$}c<{$}>{$}c<{$}}
    \toprule
    Simulation name & r_\text{Mn:\chem{^3\!He},hydro} & w_\text{ox}/(\si{\metre\per\day}) & \mathcal{X}_\text{thr}/\si{\pico\molar} & \mathcal{D}_\text{thr}/\si{\nano\molar} & R_\text{Mn:P} \\
    \midrule
    \textit{Reference}      & \num{0.10e9}               & 1\text{--}10          & 25                   & 0                      & \num{0.36e-3} \\
    \textit{NoBio}          & \num{0.10e9}               & 1\text{--}10          & 25                   & 0                      & \mathbf{0}    \\
    \textit{LowHydro}       & \mathbf{0.01\times10^9}    & \mathbf{1}            & -                    & 0                      & \num{0.36e-3} \\
    \textit{NoThreshold}    & \num{0.10e9}               & 1\text{--}10          & \mathbf{0}           & 0                      & \num{0.36e-3} \\
    \textit{OxidThreshold}  & \num{0.10e9}               & 1\text{--}10          & \mathbf{0}           & \mathbf{0.125}         & \num{0.36e-3} \\
    \bottomrule
    \addlinespace
  \end{tabular}
  \caption{List of simulations with the parameters changed compared to the reference simulation in boldface.}
  \label{tab:mang:simulations}
\end{table*}

Our second simulation is without a biological cycle of Mn (\textit{NoBio}).
In this simulation Eq.~\ref{eqn:bio} is removed from the model, or,
equivalently, $R_\text{Mn:P}$ is set to zero.
This simulation should illustrate the consequences of the (lack of) biological
incorporation and subsequent remineralisation of~Mn.

The goal of \textit{LowHydro} is to investigate whether the combination of the
high hydrothermal input of \chem{Mn_\text{diss}}, and strong aggregation
(modelled as a high settling velocity), is needed to obtain an accurate
representation of the distribution of \chem{Mn_\text{diss}}, i.e.\ one where the
predicted concentrations compare well with the observed concentrations.
Specifically, we want to explain the sharp observed Mn plumes.
To this end, we first decreased the settling velocity to a constant
\SI{1}{\metre\per\day} (or, equivalently,
$\mathcal{X}_\text{thr}\rightarrow\infty$), which, as expected, resulted in a
wide spreading of \chem{Mn_\text{diss}} and a too high \chem{[Mn_\text{diss}]}
almost everywhere in the deep ocean (not presented).
Decreasing only hydrothermal input would trivially result in a proportionally
smaller \chem{[Mn_\text{diss}]} near the vents.
For this reason \textit{LowHydro} contains two changes compared to the
\textit{Reference} simulation: a tenfold decrease in both the hydrothermal flux
and the maximum settling velocity.

In our fourth simulation, \textit{NoThreshold}, we want to see if an aggregation
threshold is needed for an accurate \chem{[Mn_\text{diss}]} simulation.
The threshold is removed by setting $\mathcal{X}_\text{thr}$ in
Eq.~\ref{eqn:settlingMnox_step} to zero.

The fifth and final simulation, \textit{OxidThreshold}, is one that has a
threshold in the oxidation process instead of the particle sinking, setting effectively
a minimum concentration of dissolved~Mn.

\subsection{Model validation}                  \label{sec:mang:observations}

\begin{figure}[ht!]
    \centering
    \includegraphics[width=\columnwidth]{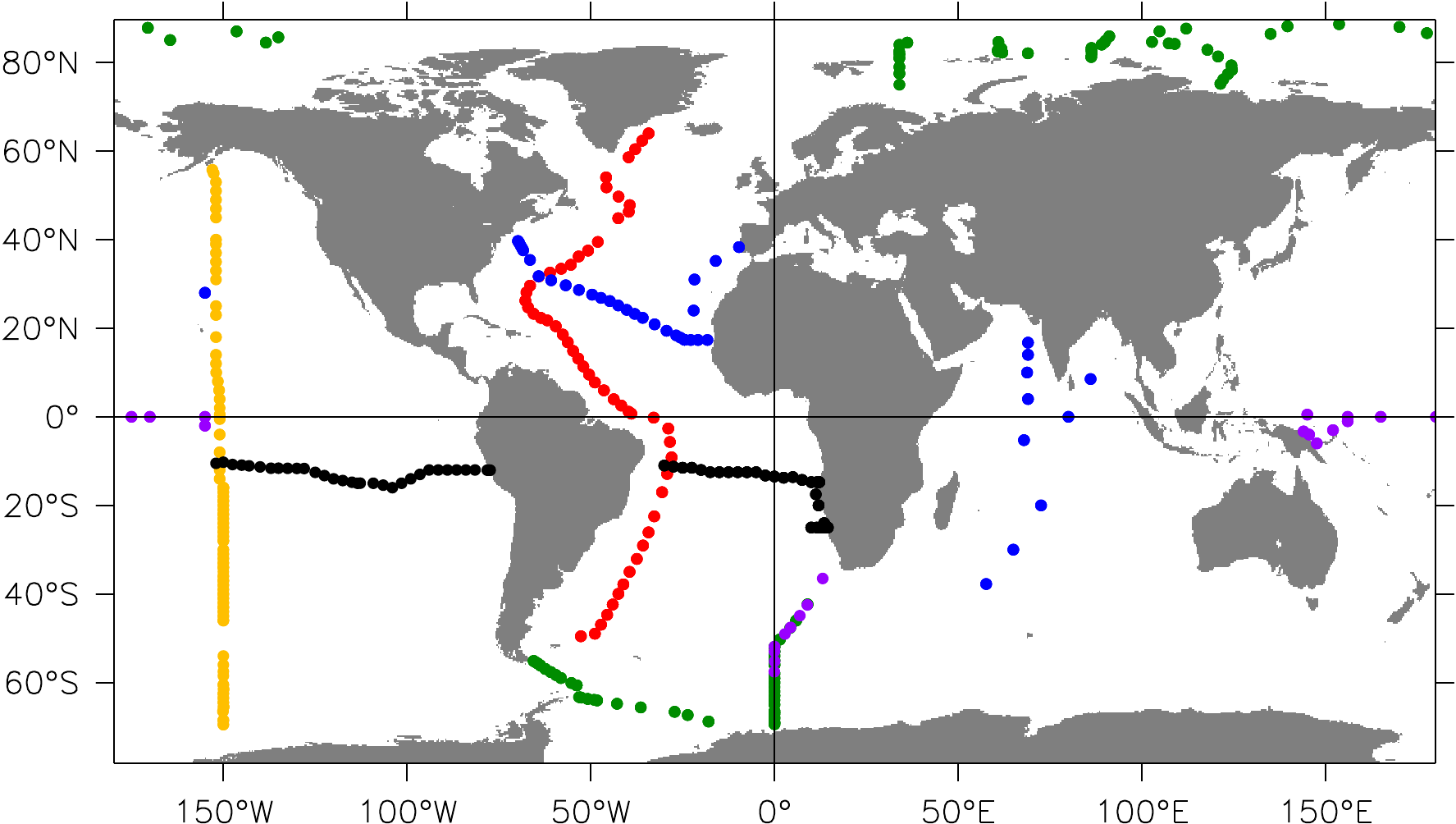}
    \caption{Transect (or expedition) names corresponding to station colours:
        {\color{forestgreen}GIPY11} in the Arctic Ocean;
        {\color{violet}GIPY4} and {\color{forestgreen}GIPY5} in the Atlantic sector of the Southern Ocean;
        {\color{blue}GI04} in the Indian Ocean;
        {\color{red}GA02} in the West Atlantic Ocean;
        {\color{blue}GA03} in the North Atlantic Ocean;
        {\color{blue}VERTEX-4} in the North Pacific Ocean;
        {\color{violet}EUCFe} in the equatorial Pacific Ocean;
        {\color{amber}CLIVAR\;P16} in the Pacific Ocean;
        {\color{black}GP16} in the South Pacific;
        {\color{black}CoFeMUG} in the South Atlantic Ocean.
        See Table~\ref{tab:data_Mn} for an overview with references
        and the number of observations.
    }
    \label{fig:cruises_Mn}
\end{figure}

In this study we mainly use data from the \textsc{Geotraces} programme, but
for a worldwide global ocean comparison one has to rely also on data
that were collected in the era before the reference samples of SAFe and
\textsc{Geotraces} were available.
Table~\ref{tab:data_Mn} lists the datasets from \textsc{Geotraces} expeditions,
as well as several other
datasets; Fig.~\ref{fig:cruises_Mn} shows the coordinates of the stations.

Most of these \href{http://www.geotraces.org/}{\textsc{Geotraces}} datasets are
part of the publicly available
\href{http://www.bodc.ac.uk/geotraces/data/idp2014/}{\textsc{Geotraces}
Intermediate Data Product 2014} (IDP) \citep{mawji2015}, except for GP16
\citep{resing2015} that will be released as part of the
\href{http://www.geotraces.org/dp/intermediate-data-product-2017}{\textsc{Geotraces}
Intermediate Data Product 2017}.

\begin{table*}[t]
  \begin{tabular}{lllllr}
    \toprule
    Transect    & Year      & Expedition            & Ocean basin           & Citation                      & \#   \\ 
    \midrule
    \multicolumn{6}{l}{\textsc{Geotraces} transects}  \\
    GIPY11      & 2007      & ARK XXII/2            & Arctic Ocean          & \citet{middag2011:mn:arctic}  &  773 \\ 
    GIPY4       & 2008      & MD166 BONUS-GoodHope  & Southern Ocean        & \citet{boye2012}              &  233 \\ 
    GIPY5       & 2008      & ANT XXIV/3            & Southern Ocean        &                               &      \\ 
                &           &                       & a) Zero Meridian      & \citet{middag2011:mn:southern} & 468 \\
                &           &                       & b) Weddell Sea        & \citet{middag2013:weddell}    &  176 \\
                &           &                       & c) Drake Passage      & \citet{middag2012}            &  221 \\
    GI04        & 2009/2010 & KH-09-5               & Indian Ocean          & \citet{vu2013}                &  233 \\ 
    GA02        & 2010      & 64 PE 319             & Northwest Atlantic Ocean  & \textit{this study}       &  384 \\ 
    GA02        & 2010      & 64 PE 321             & Northwest Atlantic Ocean  & \textit{this study}       &  504 \\ 
    GA02        & 2011      & JC\,057               & Southwest Atlantic Ocean  & \textit{this study}       &  432 \\ 
    GA03        & 2010      & US\,GT10              & North Atlantic        & \citet{wu2014}                &   91 \\ 
    GA03        & 2011      & US\,GT11              & North Atlantic        & \citet{wu2014}                &  578 \\ 
    GP16        & 2013      & US\,EPZT              & South Pacific         & \citet{resing2015}            &  874 \\ 
    \addlinespace
    \multicolumn{6}{l}{Other expeditions and datasets}  \\
    -           & 1983      & VERTEX-IV             & North Pacific Ocean   & \citet{landing1987}           &   27 \\ 
    -           & 2006      & EUCFe (\textit{Kilo Moana}) & Pacific Ocean   & \citet{slemons2010}           &  349 \\ 
    -           & 2005/2006 & CLIVAR\;P16           & Pacific Ocean         & \citet{data::milne:p16}       &  174 \\ 
    -           & 2007      & CoFeMUG               & South Atlantic        & \citet{noble2012}             &  429 \\ 
    \addlinespace
    \multicolumn{5}{l}{Total number of dissolved Mn measurements:}                                          & 5742 \\
    \bottomrule
  \end{tabular}
  \caption{Observational \chem{[Mn_\text{diss}]} used for comparison with the model simulations.
    \textsc{Geotraces} datasets are indicated by \textsc{Geotraces} transect codes.
    Their accuracies are approved by \textsc{Geotraces} on the basis of results
    of reference samples and cross-over stations.
    At the time there were no reference samples available for the other
    datasets, or they were not used for~Mn.}
  \label{tab:data_Mn}
\end{table*}

These observations are used for a visual global ocean data-model comparison.
Details on the statistical and visual model--data comparison are presented in
Appendix~\ref{sec:comparison}.
Of these observations, only the West Atlantic Ocean data (the \textsc{Geotraces} GA02
transect, consisting of 64 PE\,319,
PE\,321 and JC\,057; 1320 points) and the Zero-Meridian Southern Ocean data
(GIPY5; 468 points) have been used for statistical comparison with the model.
Moreover, only the data of the shipboard FIA have
been used, which have five stations less than the on-shore mass spectrometry
determinations (ICP-MS having 120 more measurement, thus 1440 in total).
The focus of this study is the West Atlantic Ocean for several reasons.
Firstly, recent measurements have been carried out in that region,
resulting in a large consistent (one method) dataset.
Other regions generally contain fewer measurements and are mainly based on
different methods by different analysts.
Secondly, the West Atlantic Ocean is of importance to the Atlantic
meridional overturning circulation, and hence the deep ocean cycling of for
example the major nutrients.
Therefore the West Atlantic Ocean was chosen as a key site for the
$\sim$\SI{18000}{\kilo\metre} long \textsc{Geotraces} GA02 transect for the
collection of data for dissolved Mn and a suite of other trace elements and
isotopes.
The \textsc{Geotraces} GIPY5 transect at the Zero Meridian in the Southern Ocean
complements the GA02 transect up to Antarctica, giving a more complete picture
of the ocean circulation.
For these reasons in this study we focus on the GA02 and GIPY5 transects.

The particulate \chem{[Mn_\text{ox}]} measurements from \citet{landing1987}
were used to tune the redox model.
The data from the CLIVAR\;P16 cruise are unpublished; sampling is from the
surface up to \SI{1000}{\metre} depth; its methods of analysis
are described by \citet{milne2010}.

\section{Results}

\subsection{Observations in the West Atlantic Ocean}

\begin{figure*}
    \centering
    \includegraphics[width=\linewidth]{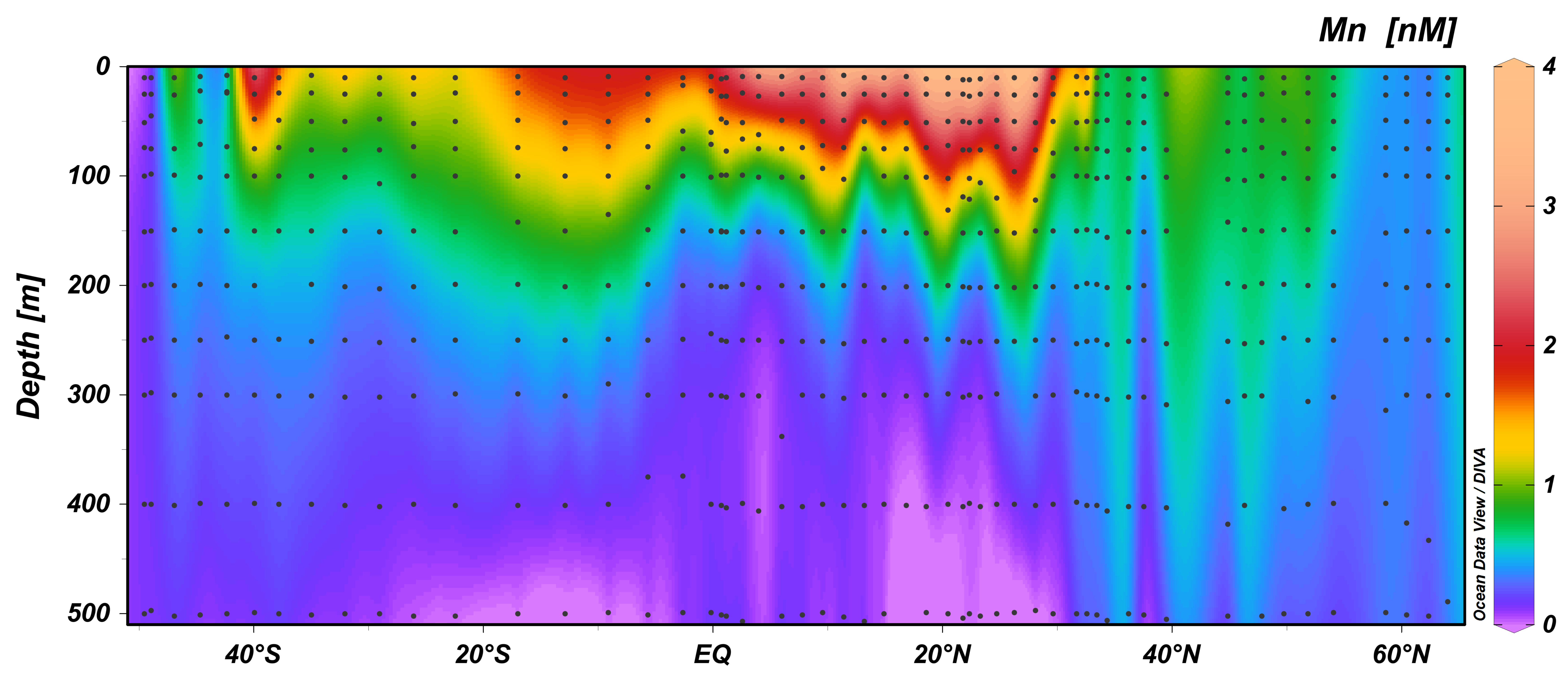}
    \caption{Observations of \chem{[Mn_{diss}]}\,(nM) along the 
            \textsc{Geotraces} GA02 transect in the West Atlantic Ocean.
            Dots are the locations of the measurements.
            Upper 500\;m.}
    \label{fig:Mn_upper}
\end{figure*}

\begin{figure*}
    \centering
    \includegraphics[width=\linewidth]{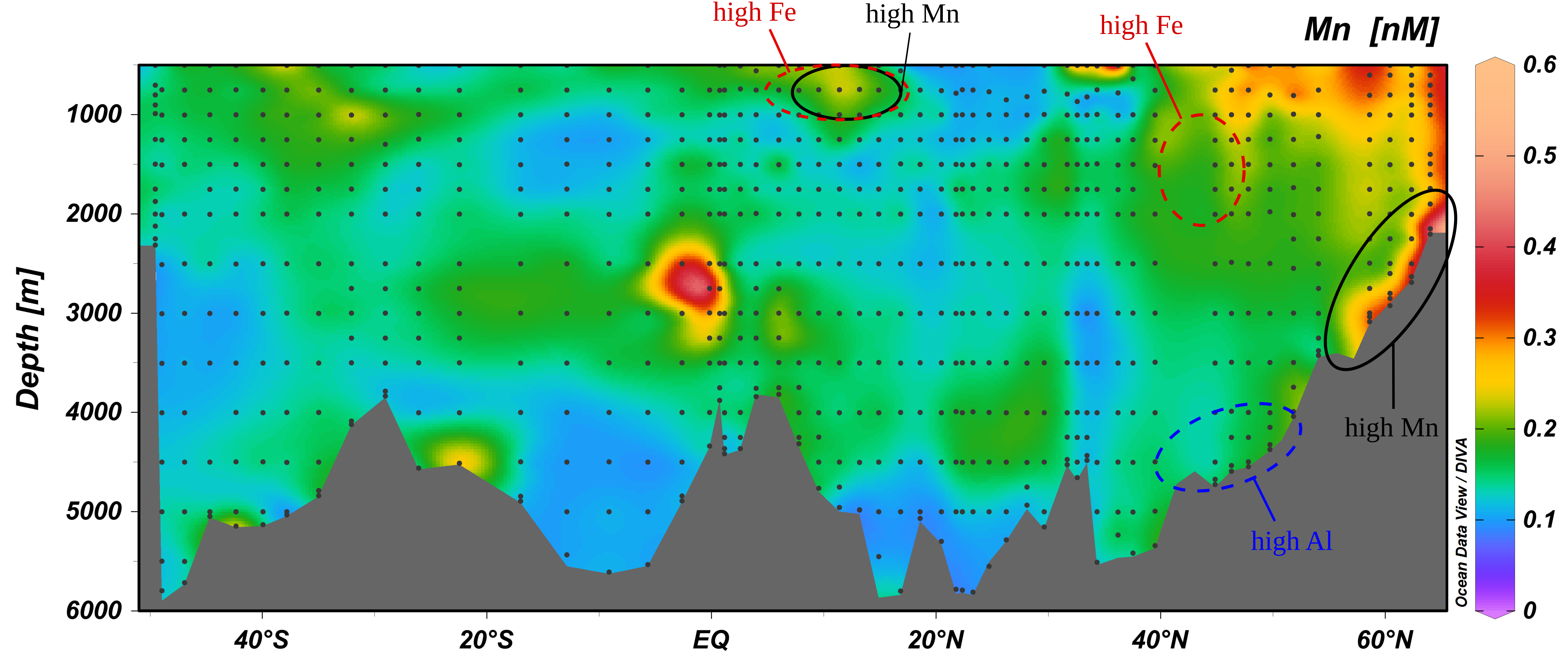}
    \caption{Observations of \chem{[Mn_{diss}]}\,(nM) along the 
            \textsc{Geotraces} GA02 transect in the West Atlantic Ocean.
            The dots denote the locations of the measurements.
            Below 500\;m.
            The red, dashed lines indicate samples with high dissolved
            iron concentrations after \citet{rijkenberg2014}.
            Similarly, the blue, dashed lines represent high dissolved aluminium
            concentrations, and are after \citet{middag2015:Al}.
            Note that the concentrations are generally much smaller than in the
            surface, and that the colour scale is different from that of
            Fig.~\ref{fig:Mn_upper}.}
    \label{fig:Mn_deep}
\end{figure*}

Figure~\ref{fig:Mn_upper} shows the West Atlantic \textsc{Geotraces} transect at
the top 500\;m.
The concentration of \chem{Mn_\text{diss}} near the ocean surface is high, going
up to \SI{4}{\nano\molar} and beyond.
The highest surface concentrations are observed in the regions of high dust
deposition as previously observed for aluminium \citep{middag2015:Al} and
consistent with dust as the main source of Mn to the surface open ocean.
The \chem{Mn_\text{diss}} distribution is mainly homogeneous in the intermediate
and deep ocean, where dissolved Mn has a concentration of about
\SIrange{0.10}{0.15}{\nano\molar}.

North of 40\degree\,N, high concentrations of dissolved Mn reach deeper than
the top couple of hundred metres, apparently because of vertical mixing.
As can be seen in Fig.~\ref{fig:Mn_deep}, some of it seems to go all the way
down to about 3000\;m.
However, other sources may be (partly) responsible as well, among which
advection from diffusive sediments and hydrothermal vents.
By plotting all deep waters at enhanced resolution of concentration
(Fig.~\ref{fig:Mn_deep}), the Mn maxima of hydrothermal plumes and the
\chem{O_2} minimum zones are better discernible.
Since Mn and Fe have a very similar redox chemistry in the oceans, not
surprisingly the maxima of Mn and Fe \citep{rijkenberg2014} generally overlap
(the red dashed and black solid ellipses in Fig.~\ref{fig:Mn_deep}).
Some of the Mn may be transported southwards by North Atlantic Deep Water (NADW),
but this cannot be deduced from this transect: already around 35\degree\,N,
\chem{[Mn_\text{diss}]} approaches near-constant deep background concentrations
such that the NADW plume is no longer discernible.
Similarly, at about 35\degree\,S, around 1000\;m depth, we observe elevated
concentrations in the northward advecting Antarctic Intermediate Water (AAIW),
but once again concentrations already reach the typical background concentration
around 20\degree\,S\@.
In the Northern Hemisphere, the subsurface waters underlying the high dust
deposition region near the equator have relatively elevated \chem{Mn_\text{diss}}
concentrations down to 750--1000\;m depth, implying some influence of the dust
deposition and particle export on the subsurface Mn concentrations, but yet
again, the typical background concentration is reached further down.
Elevated local features are located at almost \SI{2}{\kilo\metre} depth on the
Zero Meridian at 50\degree\,S \citep[][their Fig.~2; also presented as dots from
52\degree\,S on our Fig.~\ref{fig:mang:MnRef_transect}]{middag2011:mn:southern}
and almost \SI{3}{\kilo\metre} depth in the West Atlantic transect at, and just south
of, the equator, and at the Denmark Strait overflow (Fig.~\ref{fig:Mn_deep}).
The elevated Mn in the Denmark Strait Overflow Water (DSOW) for a small part coincides with
elevated aluminium attributed to sediment resuspension
\citep{vanhulten:alu_bg,middag2015:Al}.
However, there is a mismatch between the highest Al concentrations observed
around 45\degree\,N and the highest Mn in the northernmost part of the transect
\citep{middag2015:Al}.
This implies the source of Mn is related to the DSOW rather than the sediment
resuspension occurring while the DSOW advects into the North Atlantic Ocean.
The elevated deep Mn at the other locations match with elevated Fe and is most
likely of hydrothermal origin \citep{klunder2011,gerringa2015}.
Finally, at low latitudes, in the very deep ocean, \chem{[Mn_\text{diss}]}
decreases with depth, suggesting a slower circulation in that region or an
additional/enforced export of Mn to the sediment.

\subsection{Reference simulation}       \label{sec:mang:results:refsim}

After 600\;yr, both the upper 100\;m and the deep ocean relative Mn
content changed by less than 25\;ppm over a period of 10\;yr in the
\textit{Reference} simulation.
This is about a factor of four less compared to the decadal change of 100\;yr ago.
The small and reducing drift suggests that the simulation is practically in a
steady state.
Model output data are available at \url{https://doi.org/10.1594/PANGAEA.871981}.

\begin{figure*}[ht!]
    \centering
    \includegraphics[width=\linewidth]{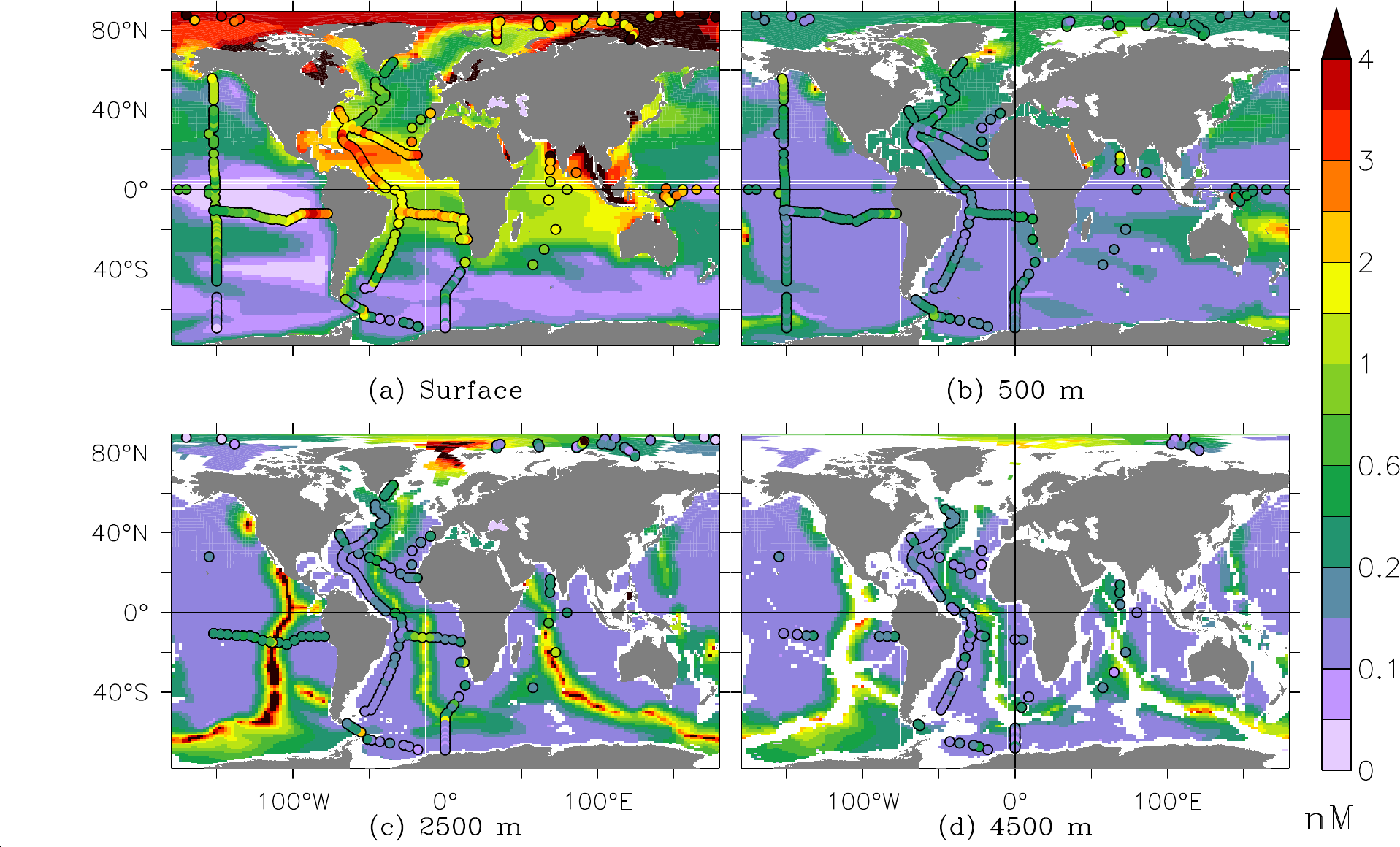}
    \caption{\chem{[Mn_\text{diss}]}\,(nM) at four depth layers in the
    world ocean for the reference simulation (\textit{Reference}) after \SI{500}{\year}
    (annual average).
    Observations are presented as coloured dots; white is the land mask of
    the model grid.}
    \label{fig:mang:MnRef_layers}
\end{figure*}

Figure~\ref{fig:mang:MnRef_layers} shows the modelled and measured dissolved Mn
concentrations at four depths; observations as coloured dots (same scale).
The dissolved Mn concentration is high in the surface of the Atlantic, Indian
and Arctic oceans, mostly consistent with the observations
(Fig.~\ref{fig:mang:MnRef_layers}a).
The model also reproduces the latitudinal gradient of \chem{[Mn_\text{diss}]} in
the Atlantic and Indian oceans, reflecting
dust deposition patterns (Fig.~\ref{fig:mang:sources}a).

\begin{figure*}[ht!]
    \centering
    \includegraphics[width=\linewidth]{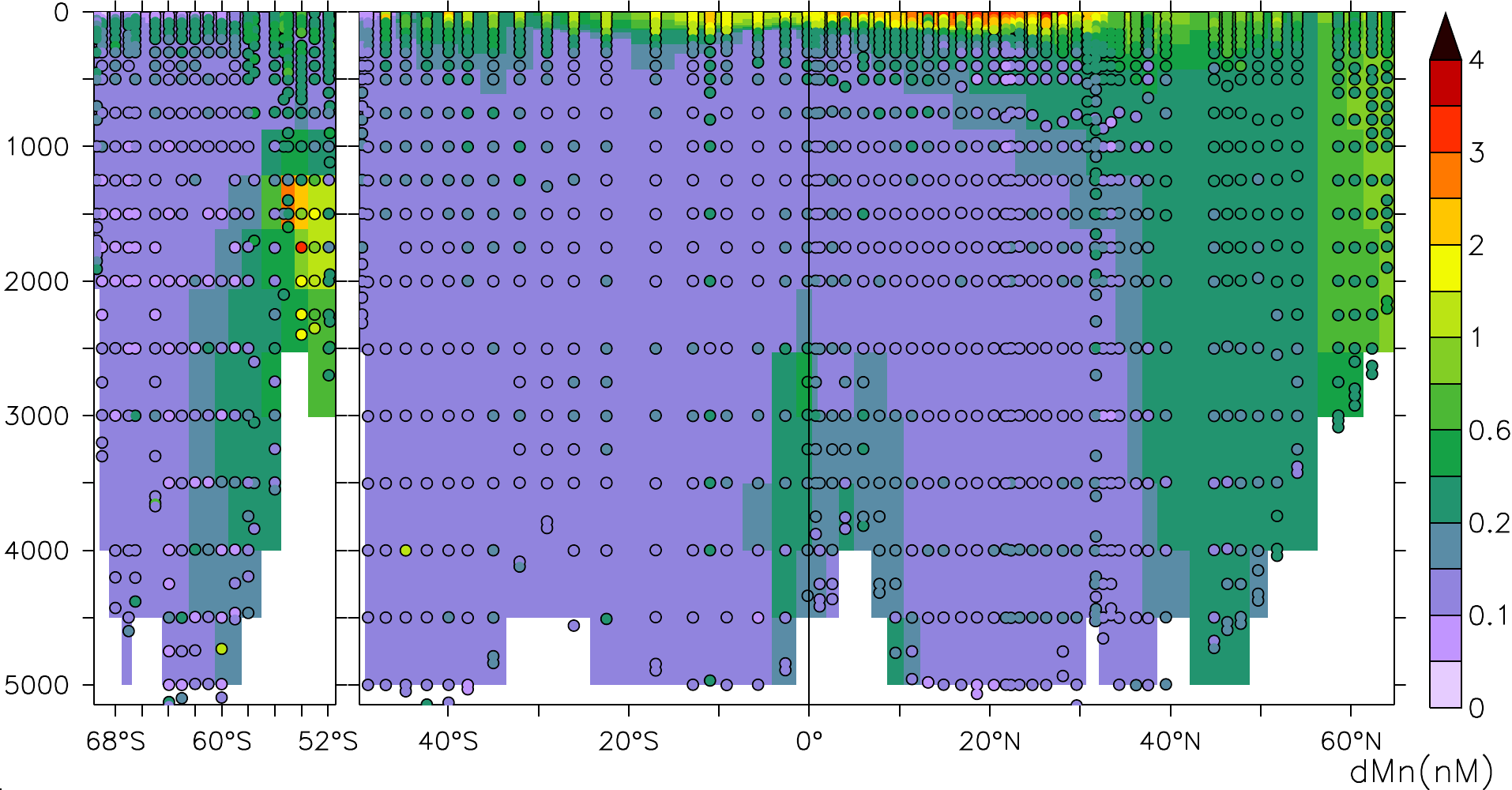}
    \caption{\chem{[Mn_\text{diss}]}\,(nM) at the Zero-Meridian section
    component of the GIPY5 dataset, and the
    West Atlantic GA02 \textsc{Geotraces} transects for \textit{Reference} after
    \SI{500}{\year} (annual average).
    Observations are presented as coloured dots.}
    \label{fig:mang:MnRef_transect}
\end{figure*}

Figure~\ref{fig:mang:MnRef_transect} shows the dissolved Mn concentrations 
at full depth in the Zero-Meridian Southern Ocean and the West Atlantic Ocean 
from the \textit{Reference} simulation; observations as coloured dots.
Lower concentrations in the deep ocean are
reproduced by the model (Figs~\ref{fig:mang:MnRef_layers}b--d
and~\ref{fig:mang:MnRef_transect}).
Both the model and observations present a mainly homogeneous distribution of
just over \SI{0.1}{\nano\molar}, though there are a number of sub-\SI{0.1}{\nano\molar}
measurements in the polar oceans.
The \chem{Mn_\text{diss}} concentrations are generally higher near the surface
compared to the deep ocean, both in the observations and the model.
This is caused by a combination of dust deposition and photoreduction.
Also the penetration of \chem{Mn_\text{diss}} from Mn-rich surface waters into the deep ocean
at around 50\degree\,N is reproduced by the model, but this is scavenged quickly
before traversing southward in the NADW\@,
which is consistent with early studies \citep[e.g.][]{bender1977}.
Finally, the \textit{Reference} simulation reproduces the measurements near
hydrothermal vents in the Atlantic and Indian oceans
(Figs~\ref{fig:mang:MnRef_layers}c and~\ref{fig:mang:MnRef_transect}).

Still, at many places, the \chem{Mn_\text{diss}} concentration is
underestimated by the model, with the most notable exceptions of the Southern
and Arctic oceans.
The underestimation is especially pronounced at the surface of the 
Pacific Ocean (Fig.~\ref{fig:mang:MnRef_layers}a).
Furthermore, in the model the Mn-rich water from the Amazon does not reach the
GA02 transect as opposed to the observations.
This potentially explains the underestimated concentration of \chem{Mn_\text{diss}}
in that region.
The simulation overestimates measured \chem{[Mn_\text{diss}]} in the Arctic
Ocean, except for some coordinates like near the Gakkel Ridge (the dark red dot
in Fig.~\ref{fig:mang:MnRef_layers}c at 90\degree\,E) \citep{middag2011:mn:arctic}.
The Gakkel Ridge is not in our hydrothermal forcing field (Fig.~\ref{fig:mang:sources}d).

\begin{figure}
    \centering
    \includegraphics[width=\linewidth]{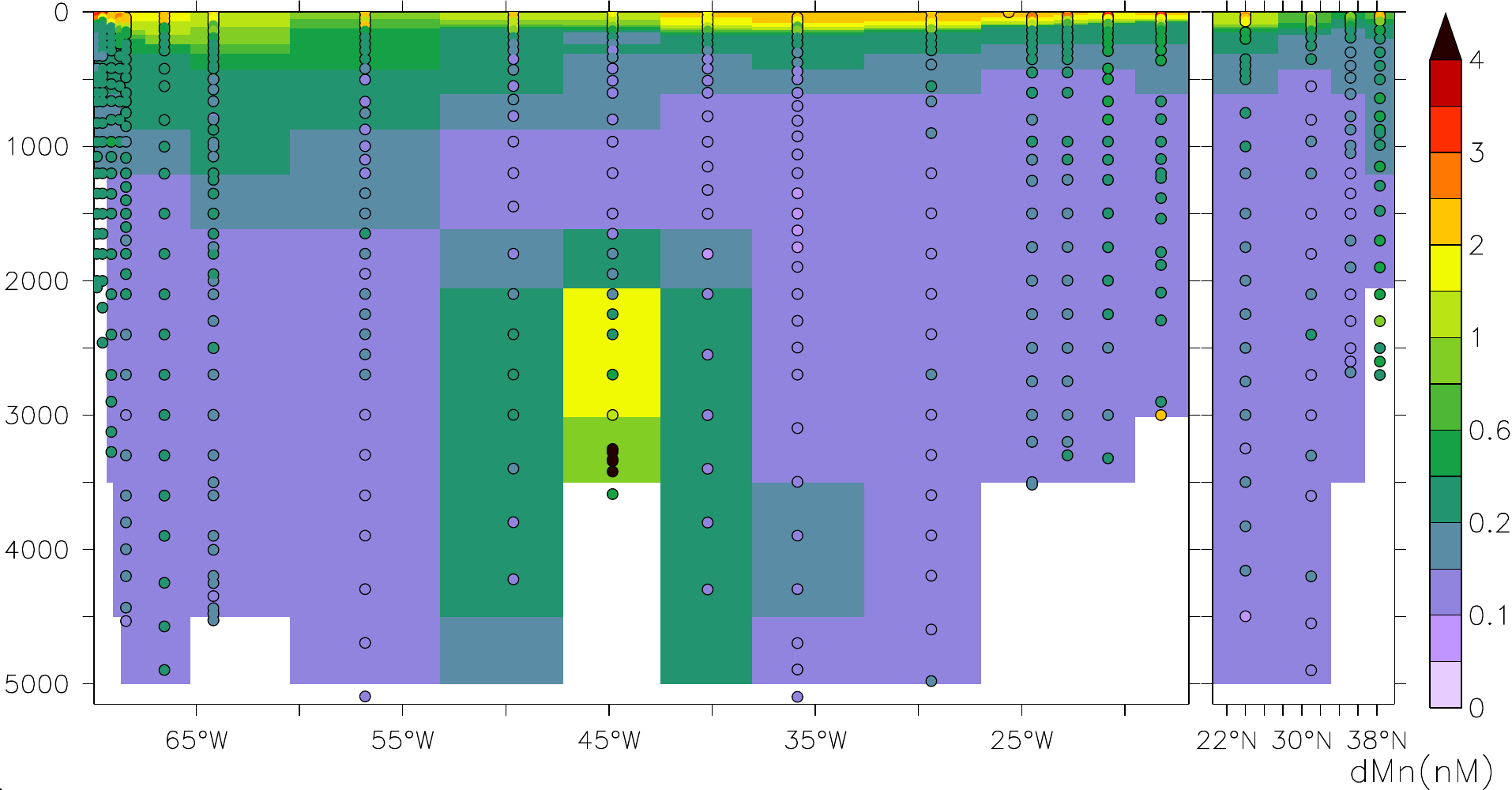}
    \caption{\chem{[Mn_\text{diss}]}\,(nM) at the North Atlantic GA03
    \textsc{Geotraces} transect for \textit{Reference} after \SI{600}{\year}
    (annual average).
    Observations are presented as coloured dots.}
    \label{fig:mang:MnRef_transect_GA03}
\end{figure}

Figure~\ref{fig:mang:MnRef_transect_GA03} presents \chem{[Mn_\text{diss}]} of
the GA03 west to east transect.
Observations from \citet{wu2014} are presented as coloured dots.
Again, the general patterns are captured, but the concentration of \chem{Mn_\text{diss}} very
close to the Mid-Atlantic Ridge is underestimated, while above the ridge (at
about 2--3\;km depth) it is overestimated.
In other words, the modelled hydrothermal \chem{Mn_\text{diss}} gradient is not as sharp as in the
observations.
It is difficult to improve this feature because of the low vertical
model resolution at that depth.

To more objectively compare the different simulations to each other, we list
several goodness-of-fit statistics in Table~\ref{tab:mang:stats_Mn}.
They compare the model simulations to observational data from the
\textsc{Geotraces} GA02 transect.
The model--data \chem{[Mn_\text{diss}]} Pearson correlation coefficient $r$ has
a value of 0.78 for the \textit{Reference} simulation.
The reliability index, the average factor by which model predictions differ
from observations, shows that on average the model differs by a factor of
1.76 from observations \citep[and Appendix~\ref{sec:comparison}]{stow2009}.

\begin{table*}
\begin{tabular}{l|lll|ll}
\toprule
Simulation          & Correlation       & RMS deviation (\si{\nano\molar})  & Reliability index & Mn (\si{\giga\mole})  & Percentage \chem{Mn_\text{ox}} \\
\midrule
\textit{Reference}  & 0.78 ($\pm 0.05$) & 0.46 ($\pm 0.05$)                 & 1.76 ($\pm 0.08$) &  440                  & 14.1 \\
\textit{NoBio}      & 0.79 ($\pm 0.05$) & 0.45 ($\pm 0.05$)                 & 1.76 ($\pm 0.08$) &  409                  & 14.0 \\
\textit{LowHydro}   & \textbf{0.64} ($\pm 0.07$) & \textbf{0.60} ($\pm 0.05$) & \textbf{2.90} ($\pm 0.08$) & 1023       & 15.8 \\
\textit{NoThreshold}& 0.78 ($\pm 0.05$) & 0.46 ($\pm 0.05$)                 & \textbf{2.03} ($\pm 0.11$)   &  373       & 13.7 \\
\textit{OxidThreshold}& 0.78 ($\pm 0.05$) & 0.46 ($\pm 0.05$)               & 1.82 ($\pm 0.09$) &  405                  & 12.6 \\
\bottomrule
\end{tabular}
\caption{Statistical model--data comparison for \chem{[Mn_\text{diss}]} at the
\textsc{Geotraces} West Atlantic GA02 transect.
The significance errors in the entries of \textit{Reference} are $\pm 2\sigma$ from
Monte Carlo samplings (Appendix~\ref{sec:comparison}).
Bold face values denote significant worsening compared with \textit{Reference};
deviations of other values from the \textit{Reference} value are insignificant.
The last two columns are the total modelled Mn in seawater, and the
Mn oxides as a portion of the total.}
\label{tab:mang:stats_Mn}
\end{table*}

\subsection{Biological cycle disabled}
                                        \label{sec:mang:results:NoBio}
As Mn plays an important role as a trace nutrient, it is likely that biology has an
impact on \chem{[Mn_\text{diss}]} at the ocean surface.
Therefore, a simulation has been performed where biological incorporation and
remineralisation of \chem{Mn_\text{diss}} was not included, henceforth named
\textit{NoBio}.

\begin{figure}
    \centering
    \includegraphics[width=\linewidth]{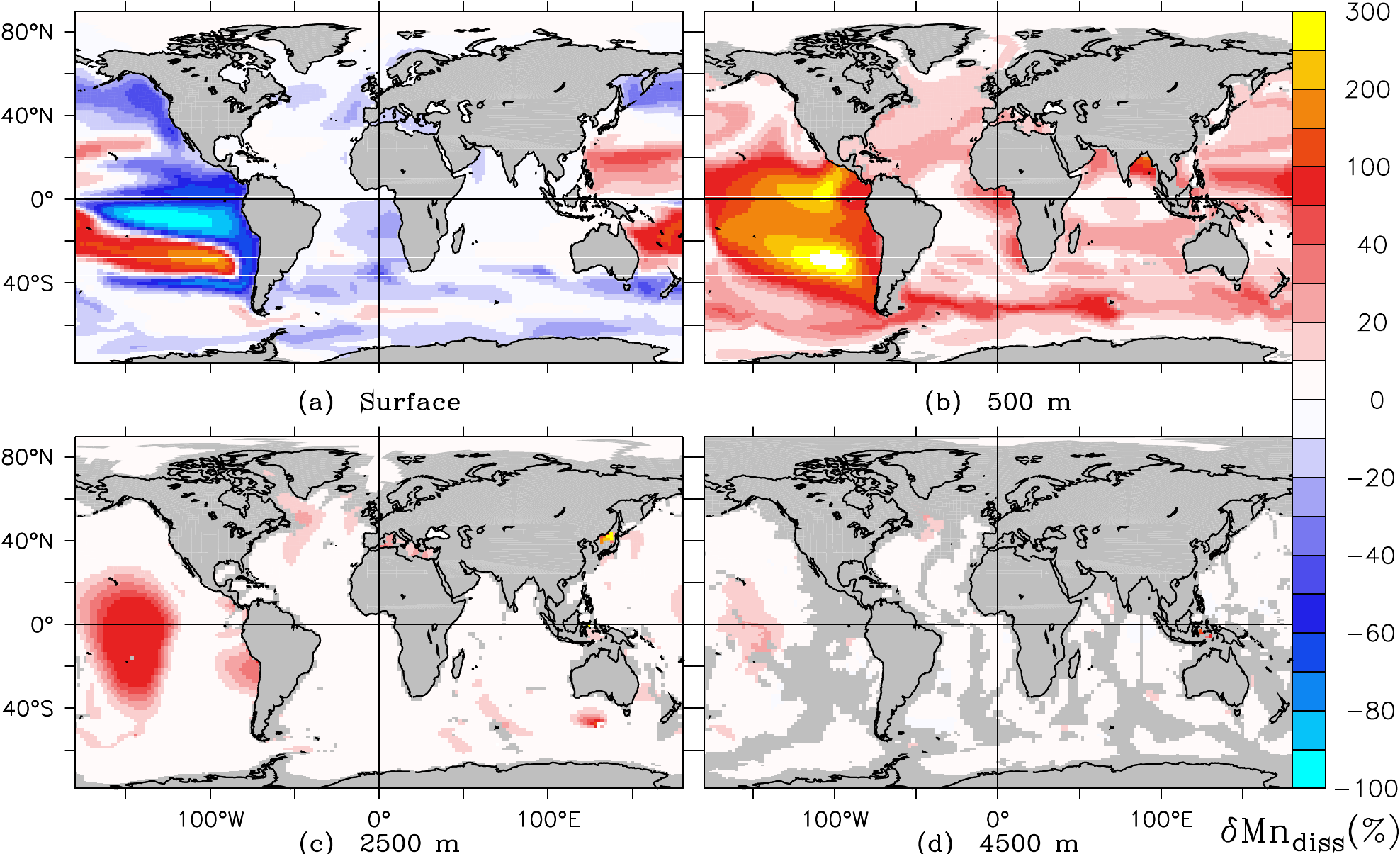}
    \caption{The effect of biological incorporation on \chem{[Mn_\text{diss}]}
at four depths: $(\textit{Reference} - \textit{NoBio})/\textit{NoBio}$.
Grey is the land mask of the model domain; the black contour is that of the real
continental coasts.}
    \label{fig:mang:NoBio_layers_reldiff}
\end{figure}

\begin{figure*}
    \centering
    \includegraphics[width=\linewidth]{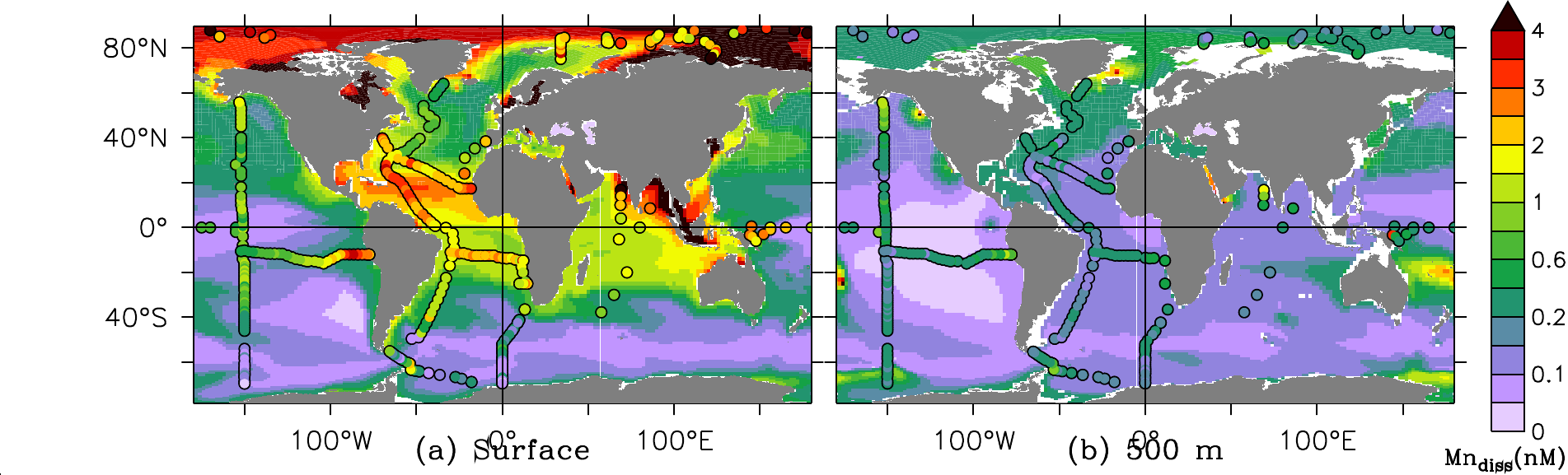}
    \caption{Annual average of \chem{[Mn_\text{diss}]}\,(nM) at two depth in the
    world ocean for the simulation without a Mn biological cycle (\textit{NoBio})
    of year 500 after forking from \textit{Reference}.}
    \label{fig:mang:NoBio_layers}
\end{figure*}

Across much of the surface
ocean, the biological cycle of Mn in \textit{Reference} causes a notable decrease
of \chem{[Mn_\text{diss}]} compared to \textit{NoBio}, though in the Pacific
Ocean there is an increase at around 20\degree\,N and especially around
20\degree\,S (Fig.~\ref{fig:mang:NoBio_layers_reldiff}).
The decrease in the Pacific Ocean is up to 100\,\% near the equator and up to
80\,\% at 40\degree\,S\@.
In the Atlantic Ocean, \chem{[Mn_\text{diss}]} has not changed much between the
two model simulations, explaining the insignificant change in the model--data
comparison (Table~\ref{tab:mang:stats_Mn}).
Deeper in the Pacific Ocean, \chem{[Mn_\text{diss}]} is higher in
\textit{Reference} compared to~\textit{NoBio}.

Figure~\ref{fig:mang:NoBio_layers} shows \chem{[Mn_\text{diss}]} of the model
compared with the observations at two depths for the \textit{NoBio} simulation.
The changes between the simulations are only big in the Pacific Ocean
(Fig.~\ref{fig:mang:NoBio_layers}).
Elsewhere biology does not appear to have a big effect on the
\chem{Mn_\text{diss}} concentration.
At the surface of the equatorial Pacific Ocean, concentrations are higher in the
\textit{NoBio} simulation.
Since in \textit{Reference} \chem{[Mn_\text{diss}]} is too low compared to the
US East Pacific Zonal Transect (EPZT) (GP16), \textit{NoBio} compares better with the observations.
Deeper in the Pacific Ocean, \textit{Reference} has higher
\chem{[Mn_\text{diss}]}, hence is better than in \textit{NoBio}.

\subsection{Hydrothermal flux and export reduced}
                                        \label{sec:mang:results:LowHydro}
We have shown that the \textit{Reference} simulation gives a adequate
representation of the effects of hydrothermal vents and the background
concentration in the deep ocean.
This is achieved by setting a large \chem{Mn_\text{diss}} flux from hydrothermal
vents and a big maximum settling velocity at the depth of the vents to remove
the hydrothermal Mn away from the source.
The simulation presented in this section, \textit{LowHydro}, is meant to test if
the high flux is necessary for an accurately modelled \chem{Mn_\text{diss}} distribution.

\begin{figure}[ht!]
    \centering
    \includegraphics[width=\linewidth]{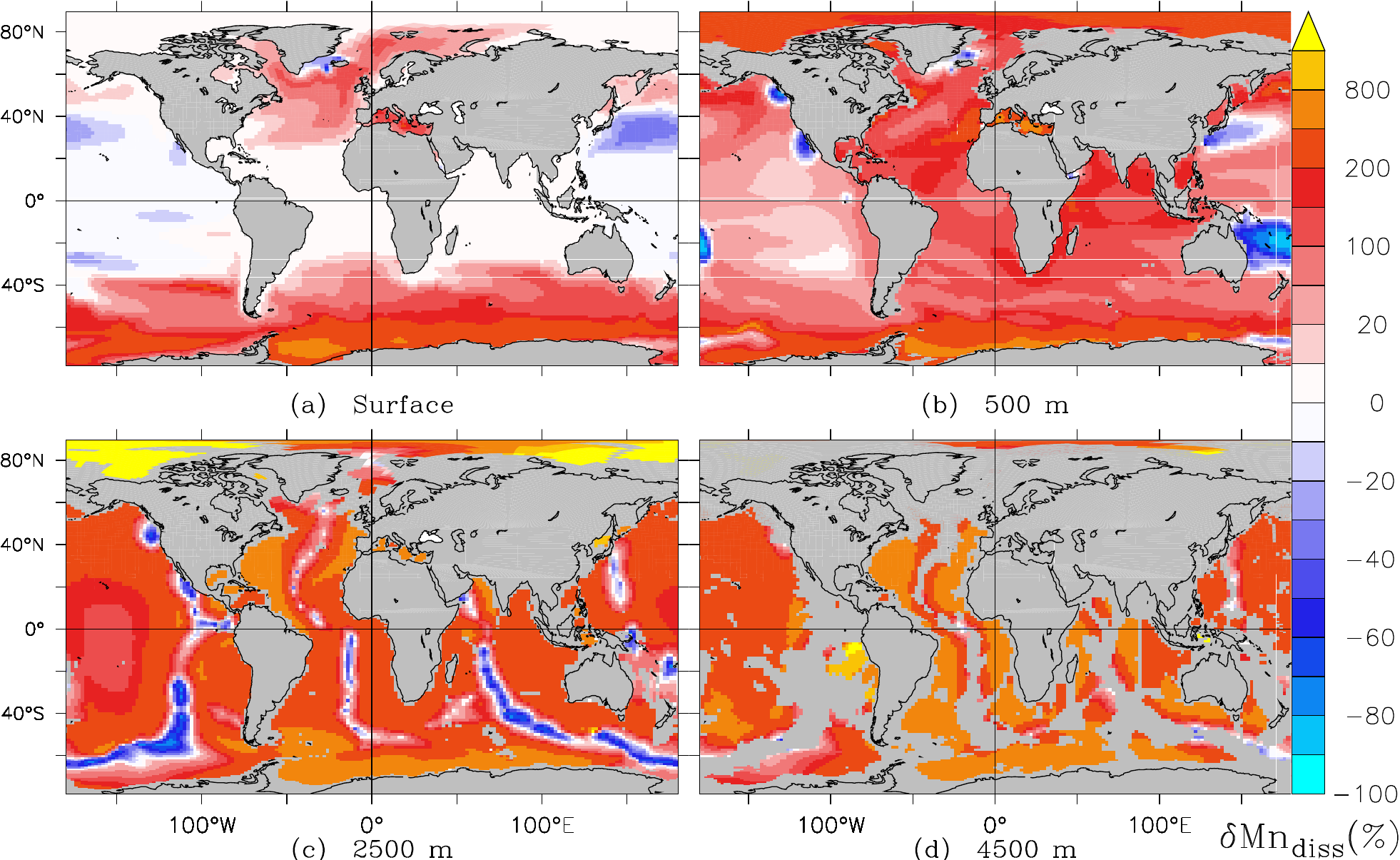}
    \caption{Result from the simulation with hydrothermal vent Mn input
    decreased and settling velocity decreased (\textit{LowHydro}).
    Relative difference in \chem{[Mn_\text{diss}]} (\%) between \textit{Reference}
    and \textit{LowHydro}: (\textit{LowHydro}--\textit{Reference}) /
    \textit{Reference}.
    To represent the changes of much larger than 100\,\%, the scale is
    increased from +50\,\% upwards.}
    \label{fig:mang:LowHydro_layers_reldiff}
\end{figure}

Figure~\ref{fig:mang:LowHydro_layers_reldiff} shows the relative change
in \chem{[Mn_\text{diss}]} at four depths when both the hydrothermal
input and the maximum settling velocity are decreased by a factor of 10
(\textit{LowHydro}).
In the surface ocean (Fig.~\ref{fig:mang:LowHydro_layers_reldiff}a)
there are both moderate increases and decreases.
However, there are very
large increases in places with deep convection that connect with the deep ocean
(over 500\,\% in the Weddell Sea).
In the deep ocean below about \SI{1}{\kilo\metre} depth, such a large
increase is not limited to the Weddell Sea but stretches over much of
the Atlantic and Indian oceans (Fig.~\ref{fig:mang:LowHydro_layers_reldiff}c and~d).
Exceptions are the locations near the oceanic ridges, where there is
hydrothermal activity.
At those locations \chem{[Mn_\text{diss}]} decreased by up to 80\,\%.

\begin{figure}[ht!]
    \centering
    \includegraphics[width=\linewidth]{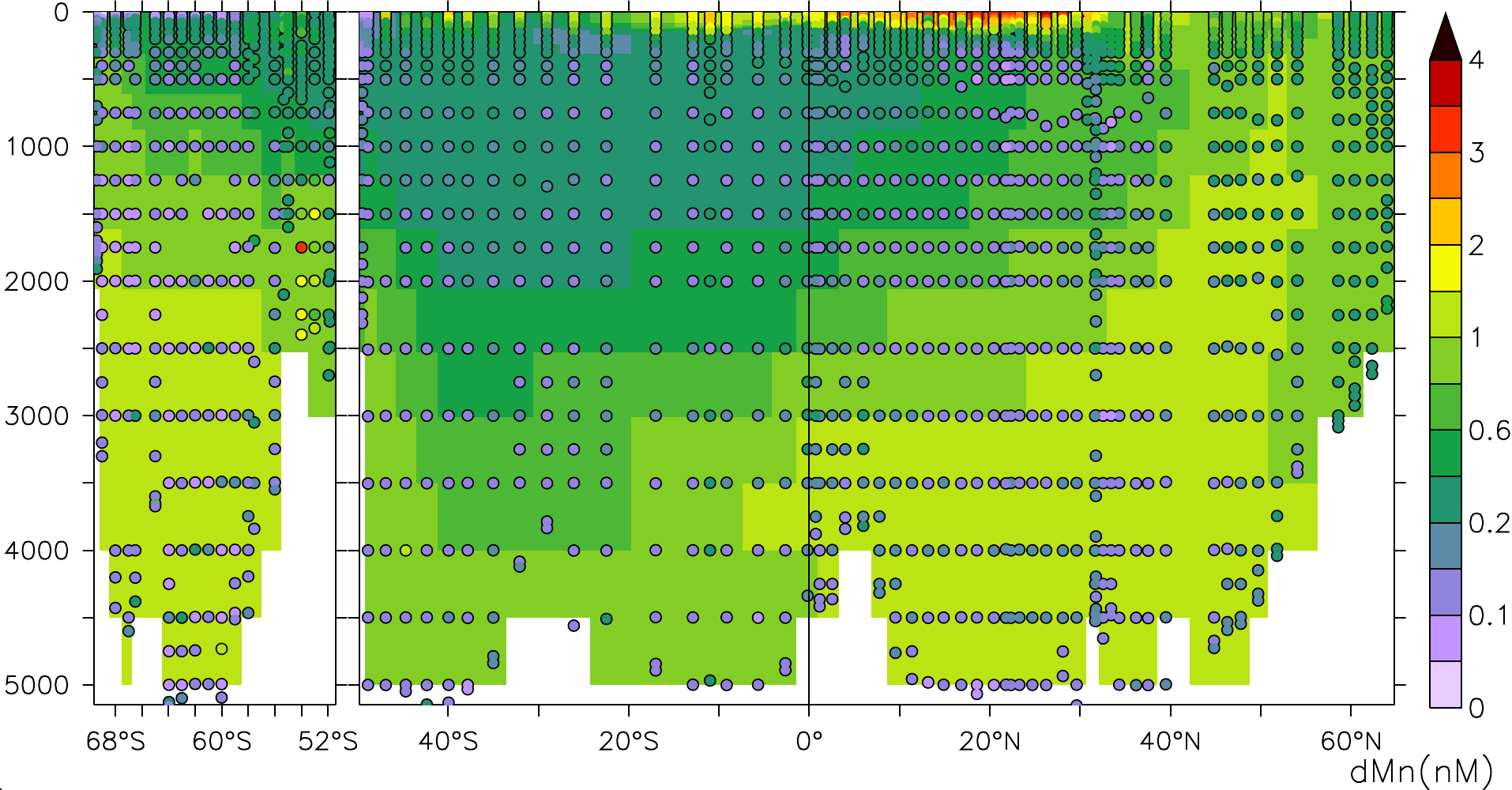}
    \caption{\chem{[Mn_\text{diss}]}\,(nM) from the simulation with
    hydrothermal vent Mn input decreased and settling velocity decreased
    (\textit{LowHydro})
    at the West Atlantic GA02 and at the Southern Ocean Zero-Meridian part of the
    GIPY5 \textsc{Geotraces} transects.}
    \label{fig:mang:LowHydro_transect}
\end{figure}

The \chem{[Mn_\text{diss}]} transects in Fig.~\ref{fig:mang:LowHydro_transect} show that
\textit{LowHydro} performs worse than the \textit{Reference} simulation.
For instance, the high \chem{[Mn_\text{diss}]} with a clear hydrothermal
origin at 54\degree\,S has disappeared, while at the same time the deep
ocean filled up with \chem{Mn_\text{diss}} resulting in a consistent
overestimation of about a factor of five in most of the deep West
Atlantic Ocean.
In the South Pacific Ocean and the Indian Ocean the signature of hydrothermal
input of \chem{Mn_\text{diss}} is -- though smaller and worse than in \textit{Reference}
-- still clearly present, with \chem{[Mn_\text{diss}]} values
near the ridges distinctly different from the ``background'' concentration
(results not presented).
Furthermore, the statistics on the points at the West Atlantic GA02
transect of \textit{LowHydro} compared to \textit{Reference} unambiguously
show that \chem{[Mn_\text{diss}]} worsened (Table~\ref{tab:mang:stats_Mn}).
The \textit{LowHydro} simulation is significantly worse in
all three statistics: the gradients from
hydrothermal vents have disappeared, and
\chem{[Mn_\text{diss}]} is much too high throughout the ocean compared to \textit{Reference}.

\subsection{Aggregation threshold disabled}   \label{sec:mang:results:NoThreshold}
\begin{figure}[ht!]
    \centering
    \includegraphics[width=\linewidth]{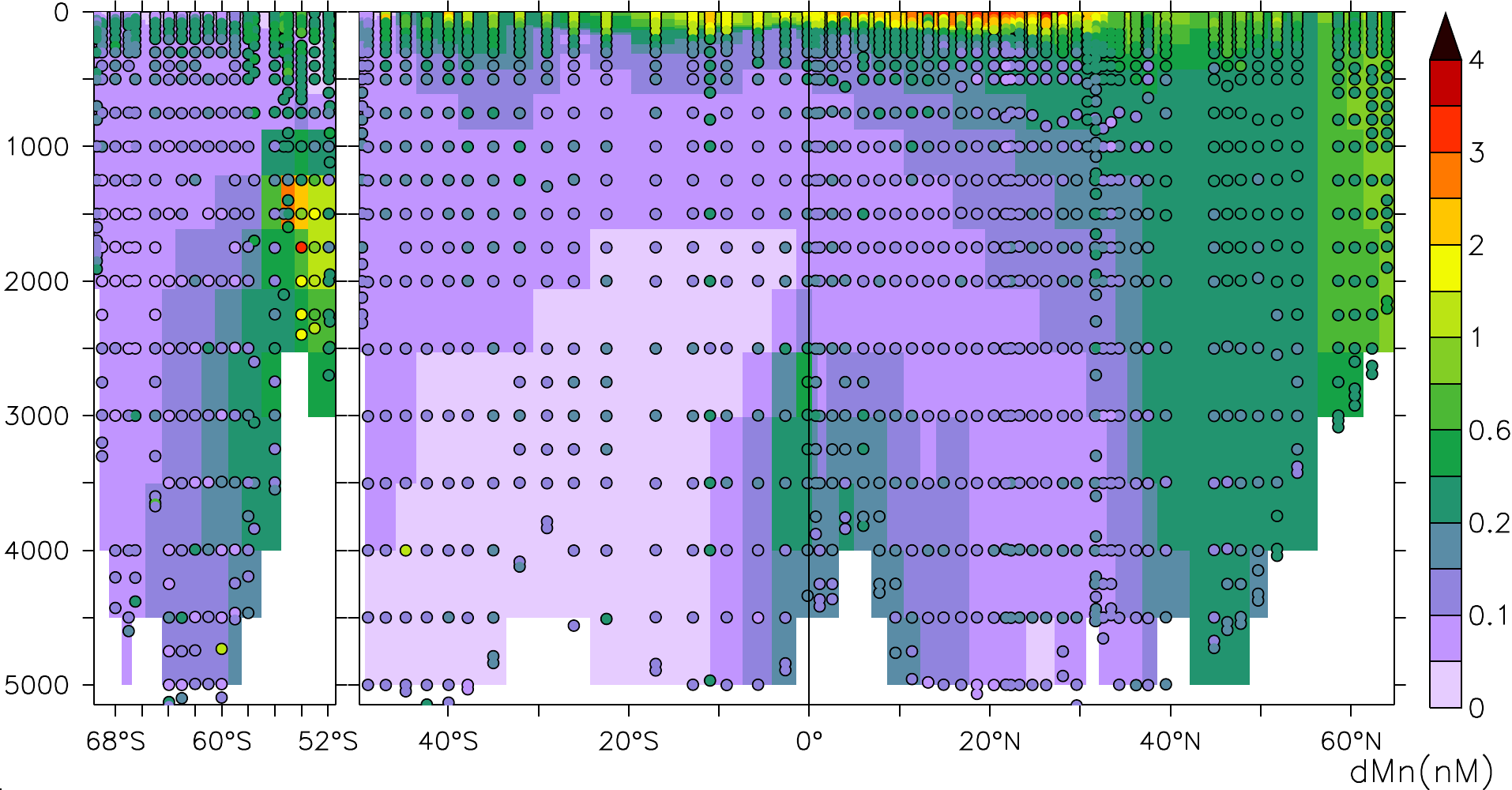}
    \caption{
    \chem{[Mn_\text{diss}]}\,(nM) at the West Atlantic GA02 and at the Southern
    Ocean Zero Meridian part of the GIPY5 \textsc{Geotraces} transects.
    Simulation without threshold (\textit{NoThreshold}).}
    \label{fig:mang:NoAggrThreshold_transect}
\end{figure}

The simulation \textit{NoThreshold} does not impose the aggregation threshold,
meaning that settling is unconstrained in this simulation.
Figure~\ref{fig:mang:NoAggrThreshold_transect} presents \chem{[Mn_\text{diss}]}
from \textit{NoThreshold} at the GA02 and GIPY5 transects.
In the intermediate and deep ocean south of 40\degree\,N the concentration of
\chem{Mn_\text{diss}} is generally more than 50\,\% smaller than that in
\textit{Reference} (Fig.~\ref{fig:mang:MnRef_transect}).

The homogeneous, already low background concentration of \chem{Mn_\text{diss}}
is reduced to close to zero (at least in the deep South Atlantic Ocean), while
the hydrothermal signals are still correctly represented in the
\textit{NoThreshold} simulation.
This explains that neither the correlation coefficient nor the root-mean-square
deviation, based on the West Atlantic GA02 data, of \chem{[Mn_\text{diss}]} in \textit{NoThreshold} differ
significantly from those of \textit{Reference}.
This means that away from ocean ridges the spatial variation of the modelled \chem{[Mn_\text{diss}]} is
similar to that of the observations \citep{stow2009}.
However, the reliability index, the average factor by which model predictions
differ from observations, here also based on the GA02 data, has changed from 1.76 to 2.03, a change that is
significant by six standard deviations (Table~\ref{tab:mang:stats_Mn}).
Therefore, the \textit{NoThreshold} simulation is much worse than the
\textit{Reference} simulation.

\subsection{Oxidation threshold enabled}   \label{sec:mang:results:OxidThreshold}
\begin{figure}[ht!]
    \centering
    \includegraphics[width=\linewidth]{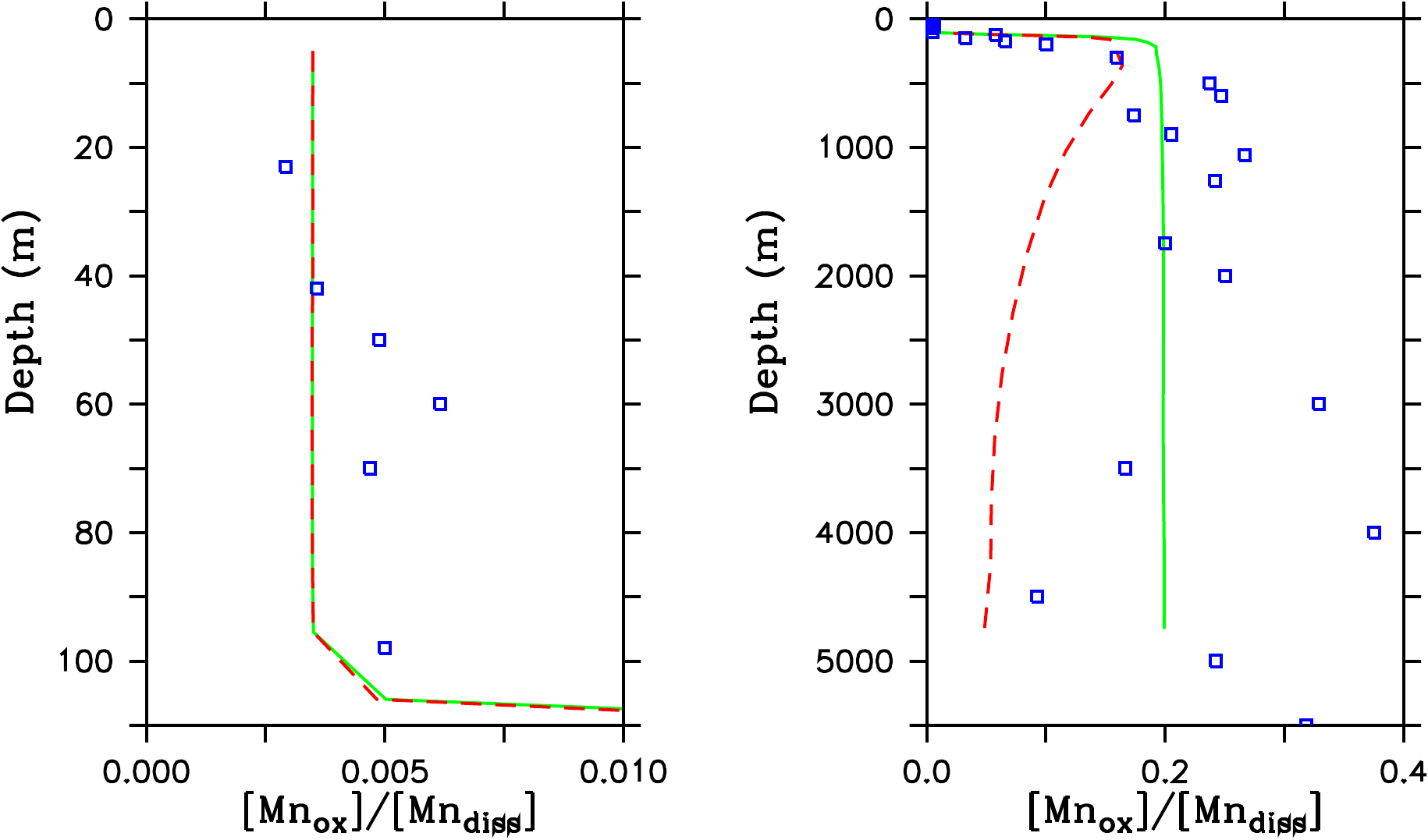}
    \caption{\chem{[Mn_\text{ox}]}/\chem{[Mn_\text{diss}]} in the
    Pacific Ocean at the VERTEX-IV station in the upper 110\;m (left)
    and at full depth (right).
    Blue squares are observations, the green line is the \textit{Reference}
    simulation, and the red dashed line is the ratio from \textit{OxidThreshold}.}
    \label{fig:oMn_dMn}
\end{figure}

As a final simulation, we limited the oxidation, rather than the settling of the
manganese oxides.
The resulting dissolved Mn concentrations are similar to those of
\textit{Reference} which has an aggregation threshold.
Figure~\ref{fig:oMn_dMn} shows the
\chem{Mn_\text{ox}}/\chem{Mn_\text{diss}} concentration ratio of the
two simulations as well as observations at (155\degree\,W, 28\degree\,N).
The blue squares represent the ratios between the particulate and dissolved Mn
measurements taken during the VERTEX-IV cruise \citep{bruland1994}.
The left panel shows that this ratio lies near 0.005 in the upper
\SI{100}{\metre}, and the right panel shows the deep ocean that has higher
values.
The green line is the \chem{Mn_\text{ox}}/\chem{Mn_\text{diss}} concentration
ratio from the \textit{Reference} simulation, and the red, dashed line is the
\textit{OxidThreshold} simulation.
For the \textit{Reference} simulation, 
the modelled $\mathcal{X}/\mathcal{D}$ generally lies at
the lower end of the observed ratio, while only between 200 and 400\;m particles
are overestimated compared to dissolved Mn.
Whereas in the upper \SI{100}{\metre} of the ocean both simulations are
consistent with those from the VERTEX-IV cruise, in the deep ocean only the
\textit{Reference} simulation compares well with the observations.
In \textit{OxidThreshold} the ratio is underestimated.

\section{Discussion}                    \label{sec:mang:discussion}

Many of the features of the \chem{Mn_\text{diss}} distribution in the
world ocean are reproduced by our \textit{Reference} simulation
(Sect.~\ref{sec:mang:results:refsim}).
However, some discussion on the assumptions of the underlying processes in the
model is required.

\subsection{Margin sediments}

We have chosen to use a simple sediment flux parameterisation for this study.
In our model, \chem{Mn_\text{diss}} is added to bottom water from anoxic
sediments analogously to the iron flux in the model \textsc{Pisces}
\citep{aumont2015}.

On the one hand, our model may underestimate the flux.
In fact, with Mn reduction a larger free energy is released compared to
Fe reduction, so that Mn oxides reduce more easily \citep{froelich1979}.
Furthermore, high benthic Mn fluxes have been observed from eastern North
Pacific marine sediments \citep{mcmanus2012}.
On the other hand, at some large-shelf regions, like in parts of the Arctic
Ocean, the Mn flux is overestimated, because the low model resolution does not
handle shelf regions well.
This problem is similar for iron \citep[their Figure~8c,\,d]{aumont2015}.

\citet{slomp1997} measured pore water concentrations of dissolved Fe and Mn.
The Mn/Fe ratios based on their pore water concentrations typically range from
0.2 to 0.5, but sometimes up to~2.5.
The reduction rate found by their reaction--diffusion model yields a Mn/Fe
reduction rate ratio of 0.02 to~0.2.
\citet{balzer1982} reports fluxes from sediment to bottom water of dissolved Mn
and Fe that yield an average ratio of~0.5.
The value that we used, 0.2, lies between the intermediate and lower
end of the reported values.
This choice is mainly due to the fact that the crude iron flux parameterisation
results in increases of the Mn flux that are out of proportion in some regions
of the ocean like the eastern Arctic Ocean and the Indonesian Throughflow.
This is a known shortcoming of this parameterisation that can also be seen in the
modelled iron distribution \citep{tagliabue2011:iron}.
In the Arctic Ocean, a higher sedimentary Mn input would induce an even stronger
overestimation than it already does with \chem{Mn/Fe}$=0.2$
(Fig.~\ref{fig:mang:sources}c).
The shortcoming had also been concluded by \citet{vanhulten:alu_jms} who presented a
sensitivity study that used the same parameterisation for aluminium sediment
input as for Mn here.

Nevertheless, a higher \chem{Mn_\text{diss}} input would be beneficial for the
simulation in the Atlantic; and especially the Pacific
Ocean, where
\chem{[Mn_\text{diss}]} is underestimated everywhere but south of
50\degree\,S\@.
This general underestimation may be due to underestimation of the enhanced
dissolved Mn as previously measured in coastal upwelling regions and their
underlying Oxygen Minimum Zones (OMZs).
Reports of elevated dissolved Mn in major Pacific coastal upwelling regions are
in the California Current of the Northeast Pacific
\citep{landing1987,chase2005}, and the Peru Humboldt Current of the Southeast
Pacific (GP16, \citet{resing2015,hawco2016}).
The mobilisation of Mn in these two Pacific upwelling systems is comparable to
the elevated dissolved Mn in two other classical upwelling systems, off Namibia
in the Southeast Atlantic (CoFeMUG, \citet{noble2012}), and in the Northwest
Indian Ocean (GI04, \citet{saager1989,vu2013}).
Indeed in the surface waters (Fig.~\ref{fig:mang:MnRef_layers}a) the model does
predict elevated dissolved Mn off California and off Peru, but simulated
concentrations in the $\sim 0.6$ to $\sim \SI{2}{\nano\molar}$ range are lower
than measured values up to 4 or even 10\;nM off California, and up to $\sim
\SI{3.4}{\nano\molar}$ off Peru (Fig.~\ref{fig:mang:MnRef_layers}a).
Otherwise off Namibia there appears to be more agreement between the simulated
and the measured dissolved Mn in surface waters.
The upwelling regions also are characterised by an extensive OMZ in the subsurface
waters.
In the eastern Pacific a tongue of very low dissolved \chem{O_2} is most apparent
at around 300\;m depth \citep[their Figure~1]{hawco2016} and would yield
elevated dissolved Mn.
In our model at the 500\;m depth horizon (Fig.~\ref{fig:mang:MnRef_layers}b)
this is indeed seen off California and off Namibia, but not reproduced very well
off Peru where some higher measured Mn up to $\sim$2\;nM was in fact observed.
The very low dissolved \chem{O_2} and high dissolved Mn in these OMZ waters,
lead to strong spatial gradients, and a more fine resolution of the model in
upwelling and OMZ regions would be required.

To summarise, we had chosen to use a simple sediment flux parameterisation,
which is not completely adequate for representing \chem{Mn_\text{diss}} fluxes
from OMZ sediments.
For future work a finer regional resolution of the model in upwelling/OMZ
regions, better parameterisation, or even an explicit sediment submodel, should
be taken into consideration.

\subsection{Biological cycle}

\subsubsection{General discussion and the GIPY5 transect}

Our model includes biological processes involving manganese in a way very similar to
that of phosphorus, but the rate variables are multiplied by a typical Mn/P ratio
based on the measured plankton content ratio.
We had chosen to follow all phosphorus cycling as described by \citet{aumont2015} for
consistency reasons.
We did not want to evaluate all potential details of biology but rather
take a simple approach.

At this stage, we lack observational constraints allowing us to develop a more
complex representation of the biological Mn cycle.
Specifically, even though in the upper waters of the Southern Ocean the uptake and
remineralisation cycle of \chem{Mn_\text{diss}} correlates well with the
nutrients \chem{PO_4^{3-}} and \chem{NO_3^-} \citep{middag2011:mn:southern}, there
is no mechanistic evidence for how the uptake-remineralisation process exactly
should work for Mn.
Of course, Mn plays an important role in biology
\citep{coale1991,middag2011:mn:southern,middag2013:weddell,browning2014}.
Whereas the trace nutrient iron often appears more significant \citep{buma1991},
manganese might even be limiting to primary production in parts of the Southern Ocean.
This is at least suggested by the observational data: at the Zero-Meridian
GIPY5 transect south of 58\degree\,S, the dissolved Mn concentration is very low at
the surface, and from $\sim$\SIrange{50}{250}{\metre} there is a maximum of
\chem{[Mn_\text{diss}]} (Fig.~\ref{fig:mang:MnRef_transect}, left transect).
This is probably due to remineralised particulate organic matter \citep{middag2011:mn:southern}.
The maximum is not found in the model.
For instance, near 66\degree\,S at \SI{1000}{\metre} upwards, the modelled
\chem{[Mn_\text{diss}]} decreases monotonically,
which suggests that biology does not explain the subsurface maximum.
However, our model may not be adequate.
Firstly, it may be unreasonable to treat Mn biologically identical to P\@.
Secondly, the subsurface maximum can be a direct consequence of a low oxygen
concentration, because that would decrease the oxidation rate~$k_\text{ox}$, and
our redox model does not depend on \chem{[O_2]}.
This is not the most probable hypothesis, because at least during the
ANT\;XXIV/3 expedition only a slight oxygen minimum occurred around 300--400\;m
depth which lies well below the Mn maximum \citep{middag2011:mn:southern}.
Thirdly, there could be problems with the underlying model.
For instance, if the vertical mixing in the underlying circulation model is too strong,
low-\chem{[Mn_\text{diss}]} waters mix from the deep Southern Ocean
upwards through the \SIrange{100}{200}{\metre} region, possibly removing
the vertical gradient of \chem{[Mn_\text{diss}]}.


The model suggests that a Mn biological cycle following P may have an important
contribution to the Southern Hemisphere, especially in the Pacific Ocean
(Fig.~\ref{fig:mang:NoBio_layers_reldiff}).
This should be tested further through field experiments and more sophisticated
model simulations.

The biologically incorporated portion of the particulate Mn fraction contributes
most to the sinking Mn flux.
The oxidised fraction has only a significant, though still smaller, contribution
near hydrothermal vents compared to incorporated Mn (not presented).
It appears that, especially for low latitudes, biological incorporation is more
important than oxidation.
\begin{figure*}
    \centering
    \includegraphics[width=\linewidth]{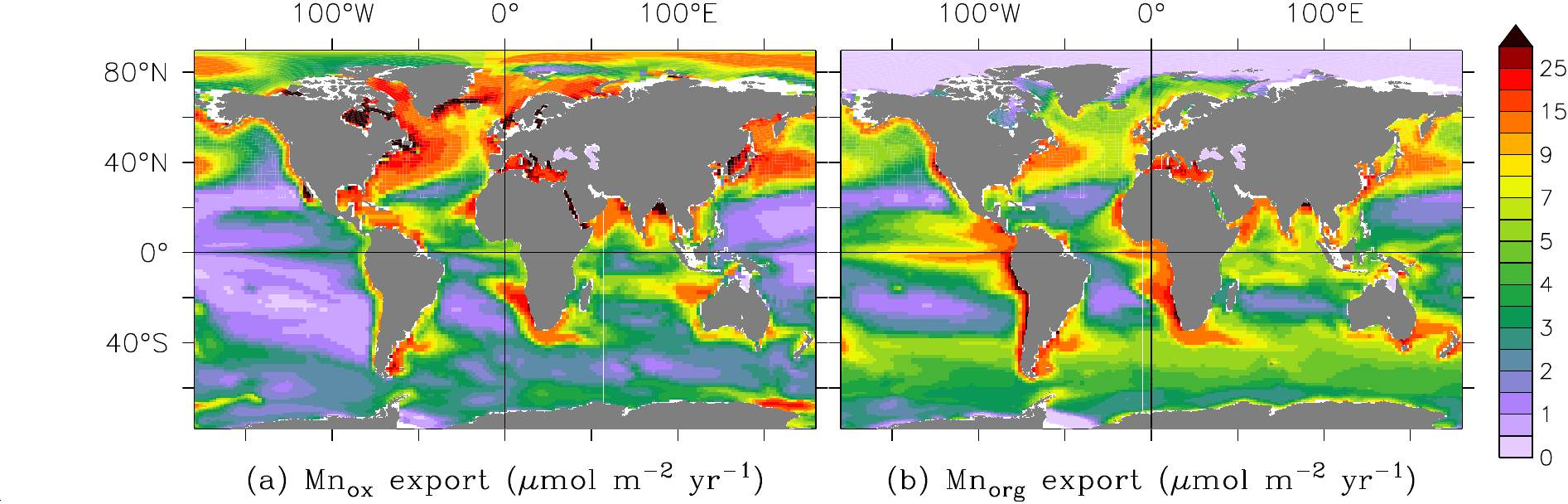}
    \caption{Manganese fluxes at 100\;m depth
    (\si{\micro\mole\per\square\metre\per\year}):
    (a) oxidised Mn, and (b) biologically incorporated~Mn.}
    \label{fig:fluxes}
\end{figure*}
The flux patterns at \SI{100}{\metre} depth for incorporated (organic) and oxidised Mn are
quite different (Fig.~\ref{fig:fluxes}).
The organic Mn flux pattern is similar to that of POC and the major nutrients,
whereas the oxidised Mn flux is typically higher, especially in the northern
seas.
However, in the low-latitude Pacific Ocean, the \chem{Mn_\text{ox}} flux is much
smaller than the organic Mn flux.
So biology strongly influences the already low \chem{[Mn_\text{diss}]} at the
surface of the Pacific Ocean.

\subsubsection{Relation to redox and settling}

Our model yields small \chem{[Mn_\text{diss}]} in surface waters of
the eastern Pacific Ocean.
Either (i) the supply of dissolved Mn from the extensive OMZ in subsurface
waters of this region \citep[their Figure~1]{hawco2016} is too low, thus not
adequately simulated by the model, or (ii) the loss due to biological uptake is
too high.
For the OMZ the observed vertical and lateral gradients of both dissolved
\chem{O_2} and dissolved Mn are very large, and would require a higher resolution
of the model grid in the OMZ regions.
This would require major revisions of the model and for the time being is merely
recommended for future work.
On the other hand, the hypothesis of perhaps too much biological removal has been
tested by simply turning off the biological module. The difference in East
Pacific surface waters (Fig.~\ref{fig:mang:NoBio_layers_reldiff}) shows a large
negative offset in exactly those surface waters that overly the strong OMZ at
300\;m depth \citep[their Figure~1]{hawco2016}.
Also in Fig.~\ref{fig:mang:NoBio_layers} we see that a model run without
biological incorporation shows surface water Mn off California and off Peru that
tend to better agree with the observations, than in the model
(Fig.~\ref{fig:mang:MnRef_layers}), and also again the lateral distribution
mimics the distribution of the underlying OMZ subsurface waters.
Thus overall it appears that the discrepancy is more due to inadequate
simulation of the OMZ regions, than due to too strong removal by the simulated
biology.

Overall, the amount of manganese (\chem{Mn_\text{diss}+Mn_\text{ox}}) in the
world ocean is slightly larger with biology (\SI{440}{\giga\mole}) compared to
without (\SI{409}{\giga\mole}, Table~\ref{tab:mang:stats_Mn}).
This is not expected, because biological incorporation would give an extra sink
of Mn incorporated in particulate organic matter.
The explanation of this paradox lies in the interaction between the biological
cycle and the settling process of \chem{Mn_\text{ox}} in the model.

We apply a concentration-dependent sinking speed, which generates an
approximately homogeneous distribution of dissolved manganese of around
\SI{0.125}{\nano\molar} in much of the deep ocean.
There is a finite range of input fluxes where this works.
If the input flux gets above a certain value, \chem{[Mn_\text{diss}]} exceeds
this background value, as it occurs in the Atlantic Ocean north of
35\degree\,N\@.
However, if the input flux is low enough, \chem{[Mn_\text{diss}]} decreases
below this background value because of slow \chem{Mn_\text{ox}} settling, like
it does in the East Pacific Ocean.
Inflow of higher \chem{[Mn_\text{diss}]} waters by large-scale circulation also
counts as a source that can elevate Mn, but the currents in the deep Pacific
Ocean are slow.

The biological processes modify this behaviour.
In \textsc{Pisces} there are two types of particulate carbon: one of large
particles that sinks with \SI{50}{\metre\per\day} (in our version) and one of
small particles that sinks with \SI{2}{\metre\per\day}.
Most carbon, thus also manganese, is present in the pool of small particles.
Even though, at least in the model, the Mn sources in the Pacific Ocean are small,
there is a lot of \chem{Mn_\text{diss}} in the surface ocean that would ultimately be
\chem{Mn_\text{ox}} if it were mixed down below the photic zone
depth.
However, with the biological cycle enabled, phytoplankton incorporates a part of
the \chem{Mn_\text{diss}}.
Particulate organic Mn (just like POC) sinks down into the deep ocean.
During the downward propagation of the organic Mn particles, they remineralise,
releasing \chem{Mn_\text{diss}} in the deep ocean.
A part of this \chem{Mn_\text{diss}} is converted to \chem{Mn_\text{ox}} that
sinks down inefficiently (\SI{1}{\metre\per\day}) when \chem{Mn_\text{ox}} has
not yet reached the equilibrium concentration $\mathcal{X}_\text{thr}$.
The remineralised Mn fills up, through oxidation, the
\chem{Mn_\text{ox}} pool in the deep ocean.

Biological processes store Mn in plankton, which is preserved from strong
Mn oxide settling (in the model).
This new pool of manganese fills the pools of dissolved and oxidised Mn in
the deep ocean.
This new pool is provided by the plankton that took it from the
\chem{Mn_\text{diss}} pool in the photic zone.
Without biology this would be available for oxidation to \chem{Mn_\text{ox}}
during night. 
Adding this extra Mn to the integrated total yields a higher Mn
quantity in \textit{Reference} than in \textit{NoBio}.

\subsubsection{Atlantic--Pacific contrast}
We saw a strong relative effect of biological incorporation in the Pacific Ocean
but not in the Atlantic Ocean.
This is because of the fact that \chem{[Mn_\text{diss}]} is already quite low in
the Pacific Ocean.
The added modelled sink of biological incorporation in the surface ocean, simply
following P uptake, quickly drains \chem{Mn_\text{diss}} concentrations towards
zero.
In reality, \chem{[Mn_\text{diss}]} is still quite high in the surface of the
Pacific Ocean.
The high observed surface concentration is primarily caused by strong Mn sources
from anoxic sediments in the Southeast Pacific Ocean, in combination with low
\chem{[O_2]} in juxtaposed waters inhibiting manganese oxidation.
The discrepancy of the model versus the observations of \chem{[Mn_\text{diss}]}
is primarily due to the underestimation of those sources, and because in the
model oxidation does not explicitly depend on \chem{[O_2]}.
At the western boundary of the Atlantic Ocean the Mn source is higher, but one
would actually expect a high Mn source at eastern boundaries where upwelling
occurs.

As dust deposition and dissolution are uncertain, we can also not exclude the
possibility that we underestimate the Mn dust flux into the (South) Pacific Ocean.
A preliminary simulation with the \citet{mahowald1999} dust flux that is
especially higher in the Pacific Ocean, shows more realistic
\chem{[Mn_\text{diss}]} throughout the Pacific Ocean.
Nonetheless, we decided to use the flux from \citet{hauglustaine2004} that is,
supposedly, more state-of-the-art.
Furthermore, the \citet{mahowald1999} dust flux did not fully solve the issue of
the low \chem{Mn_\text{diss}} concentrations near the upwelling regions.

\subsection{Redox rates}

Although we chose to model \chem{Mn} reduction and oxidation as
first-order-reaction kinetics, redox of \chem{Mn} within the water column is a
combination of several processes.
Firstly, \chem{Mn_\text{ox}} is subject to non-biological reduction,
significantly stimulated by sunlight \citep[e.g.][]{sunda1994}.
This is taken into account by using different $k_\text{red}$ for the
euphotic and aphotic zones of the ocean.
Secondly, the rate of oxidation is enhanced by microbes (bacteria and
fungi) in regions where \chem{Mn_\text{diss}} supply is high
\citep{sunda1994,tebo2005}, and it depends on the \chem{O_2}
concentration and pH\@.
This has been observed in the North Pacific Ocean \citep[and references
therein]{johnson1996}, as well as in some \textsc{Geotraces} transects in the
Atlantic Ocean.
Measurements from the West Atlantic Ocean and Zero-Meridian Southern Ocean
\textsc{Geotraces} cruises show some variations with depth, e.g.\ near
10\degree\,N and near 40--47\degree\,N (top of Fig.~\ref{fig:Mn_deep}).
This corresponds to the maximum of dissolved Fe associated with the oxygen
minimum as reported for the same cruise by \citet{rijkenberg2014}.
Our model does not reproduce such a maximum of dissolved Mn, because in the
model the Mn oxidation rate does not depend on \chem{[O_2]}.
Furthermore, fluxes from reducing sediments are not expected to reproduce the
feature either \citep{johnson1996}.
This means that at least for those regions it would be logical to include a
dependence on \chem{[O_2]}.
However, this is not the pattern that attracts the most attention when
looking at the full transect.
The striking patterns are rather the high concentration at the
surface and near the equator around \SIrange{2.5}{3}{\kilo\metre} depth
(Fig.~\ref{fig:Mn_deep}).
For this reason we have not included a dependency on \chem{[O_2]} to the
model.
In other words, while $k_\text{ox}$ depends on oxygen, we assumed here that the
oxygen concentration is generally so high that its variation does not
affect~$k_\text{ox}$.

In our model, oxidation takes place everywhere with the same
$k_\text{ox}=\SI{0.341e-3}{\per\hour}$.
Based on the underestimation of the concentrations in the surface of the Atlantic and
Pacific oceans, one may think that \chem{[Mn_\text{ox}]} is too high compared to \chem{[Mn_\text{diss}]}.
In other words, assuming approximate equilibrium, the ratio
$k_\text{ox}/k_\text{red}$ could be too high.
However, the $k$ values in our model are chosen to get values consistent with the
ratio of concentration measurements of \chem{Mn_\text{ox}} and
\chem{Mn_\text{diss}}.
To this end we set $k_\text{red,light}$ to the mean value found by
\citet{sunda1994}, then used the dissolved and particulate profiles in Fig.~4
of \citet{bruland1994} to derive $k_\text{ox}$ and~$k_\text{red,dark}$.
If we were to ignore that ratio, and increase the first-order rate constants for
reduction with a factor of five (a factor of three higher than the upper value
of the range found by \citet{sunda1994}), the model would still not yield sufficiently
high \chem{[Mn_\text{diss}]} (results not presented).

Still, one may introduce a threshold on the oxidation process (instead of the
settling of the particles).
The purpose would be to find out if this would result in higher surface
\chem{[Mn_\text{diss}]}, without having a significant effect on the ratio
between the dissolved manganese and the oxides.
In such a model simulation, oxidation only takes place when
\chem{[Mn_\text{diss}]} is higher than a certain threshold
$\mathcal{D}_\text{thr}$, equal to \SI{0.125}{\nano\molar} in
\textit{OxidThreshold}.
While this yields a distribution of \chem{Mn_\text{diss}} not
significantly worse in the West Atlantic Ocean than that in \textit{Reference}, the
\chem{Mn_\text{ox}}/\chem{Mn_\text{diss}} concentration ratio around
\SI{500}{\metre} and below is then strongly underestimated compared to the
VERTEX data (Fig.~\ref{fig:oMn_dMn}, right panel, red dashed line).
While \chem{[Mn_\text{diss}]} stays at realistic values, \chem{[Mn_\text{ox}]}
does not and approaches zero.
This pleads for using an aggregation rather than an oxidation threshold.

\subsection{Export dynamics}

The very homogeneous low dissolved concentration of Mn at $\sim$\SI{0.1}{\nano\molar}
in the deep waters of the West Atlantic GA02
section is remarkable (Fig.~\ref{fig:mang:MnRef_layers}).
Such deep water values of $\sim$\SI{0.1}{\nano\molar} were also found in the
deep waters of the Antarctic Ocean
\citep{middag2011:mn:southern,middag2012,middag2013:weddell}, as well as deep
waters of the Mediterranean Sea (2013 cruise, data will be included in the
\href{http://www.geotraces.org/dp/intermediate-data-product-2017}{\textsc{Geotraces}
Intermediate Data Product 2017}).
For the Atlantic Ocean, these deep concentrations are much lower than the
dissolved Mn values of $\sim$\SIrange{0.5}{0.6}{\nano\molar} reported for 
deep samples at several Atlantic stations in the 1990s
\citep{saager1997,statham1986,statham1998}.
The sampling resolution is too coarse to exclude spatial variation,
but it appears the values produced by pioneering efforts to study the oceanic Mn
distributions are overestimations.

The new ultraclean sampling methods \citep{rijkenberg2015}, rigorous
calibrations and excellent accuracy at the Bermuda crossover station for twelve
trace metals \citep{middag2015:BATS} now yield a much larger database of uniformly
very low Mn $\sim$\SI{0.1}{\nano\molar} in deep waters.
Given these very uniform and much lower background Mn concentrations, the
hydrothermal Mn plumes are better discernible and much larger widespread than
realised in the 1990s.
Exactly over or very near the ridge crest this is recently shown dramatically in
a plume of more than \SI{500}{\kilo\metre} wide over the Gakkel Ridge in the
high-latitude Arctic
Ocean \citep{middag2011:mn:arctic}, and in the Antarctic Ocean with a plume that
extends more than 1000 km, at a site near the Bouvet triple junction where three
ocean ridges meet \citep{middag2011:mn:southern}.
Similar long range extent of the Mn hydrothermal plume has recently been
discerned in the Pacific Ocean \citep{resing2015}.

Although the West Atlantic GA02 section is quite far west from the Mid-Atlantic
Ridge, the impact of the hydrothermal plume is still visible.
Dissolved Mn reaches a maximum of \SIrange{0.2}{0.3}{\nano\molar} in the
\SIrange{2500}{3000}{\metre} depth range just south of the equator.
This is consistent with observations along two zonal (east-west) sections across
the Mid-Atlantic Ridge.
These are (i) the part of the \textsc{Geotraces} GA03 section that passes over
the TAG hydrothermal site (USGT11-16 at 26.14\degree\,N, 44.83\degree\,W)
\citep{wu2014}, and (ii) at $\sim$12\degree\,S in the South Atlantic Ocean
\citep[their Fig.~1]{noble2012}.
In both sections a plume of dissolved Mn is visible that extends for
\SI{500}{\kilo\metre} in both zonal directions.

Remarkably, the background concentration of dissolved Mn is quite uniform in the
0.10 to 0.15\;nM range.
Preliminary equilibrium calculations predict an equilibrium of 0.19\;nM total
dissolved Mn with the solid phase pyrochroite \chem{Mn(OH)_2\,(\mathit{s})}.
This total dissolved equilibrium concentration is quite close to the observed
0.10 to 0.15\;nM background range measured in the oceans.
Please notice that in pyrochroite the Mn is divalent Mn(II) just as the dissolved Mn forms.
Disregarding organic complexes, the latter dissolved equilibrium Mn(II) forms
would comprise 0.148\;nM (78\,\%) free \chem{Mn^{2+}} ion, 0.027\;nM ($\sim
14\,\%$) \chem{MnCl^+} ion, 0.011\;nM ($\sim 5.8\,\%$) \chem{MnCl_2^0({\it aq})}
and a very low abundance 0.000028\;nM ($\sim$\,0.0015\,\%) \chem{MnOH^+} ion.
The above most simple Reaction~\ref{rcn:redox} for oxidation removal from
seawater, may perhaps be envisioned to be split into several reaction steps as
follows:
\begin{reaction}
  \begin{multlined}
    \chem{Mn^{2+}}
    \hspace{3mm}\underset{\text{equil.}}{\leftrightharpoons}\hspace{3mm}
    \chem{ MnOH^+\,(\mathit{aq})}         \hspace{2mm}
    \underset{\text{slow}}{\longrightarrow} \hspace{2mm}
    \chem{ Mn(OH)_2\,(\mathit{s})}          \\
    \hspace{10mm}\underset{\text{fast}}{\longrightarrow} \hspace{2mm}
    \chem{ Mn(III, IV) \text{solid state(s)} } \,,
  \end{multlined}
    \label{rcn:pyrochroite}
\end{reaction}
where the dominant (78\,\%) dissolved \chem{Mn^{2+}} species is via the minor
dissolved \chem{MnOH^+} species converted to solid \chem{Mn(OH)_2}, and next
this is oxidised to one or several of various oxide forms, for example
hausmannite \chem{Mn_3O_4} that is a mixed valency state (III, IV) Mn-oxide, or
\chem{Mn(IV)O_2} pyrolusite or birnessite.
Please notice that in the \emph{Reference} simulation an aggregation threshold for
particulate Mn at 25\;pM was invoked, that via redox rate constants is based on
typical background dissolved Mn concentrations that was chosen at 0.125\;nM.
The latter threshold value of 0.125\;nM was required to have the model simulate the
observed data, and is quite close to the equilibrium concentration of 0.19\;nM versus pyrochroite.
The notion of perhaps an equilibrium control is not new.
Previous pioneering measurements led to an apparent background concentration of
Mn of 0.4\;nM or higher, that is now shown to be too high, yet otherwise
attempts were made to compare these values in terms of equilibrium versus
hausmannite \chem{Mn_3O_4} \citep{klinkhammer1980}.
\citet{vonlangen1997} did propose a reaction mechanism also involving conversion
of \chem{Mn^{2+}} via \chem{Mn(OH)^+} as in above (\ref{rcn:pyrochroite}).
Further research is needed to verify and confirm the here discussed hypothesis
of equilibrium of dissolved Mn at $\sim 0.19$\;nM versus pyrochroite. 

In the \textit{NoThreshold} simulation
\chem{[Mn_\text{diss}]} is underestimated in much of the ocean.
When removing the threshold,
the RMSD significantly worsens: it appears that the threshold
is needed to reasonably simulate \chem{[Mn_\text{diss}]}.
But there are several alternatives to the aggregation threshold that may keep
\chem{Mn_\text{diss}} at a relatively homogeneous concentration.

Firstly, instead of an aggregation threshold, a criterion on oxidation
may be imposed, because, hypothetically, there may be a lower limit below which
oxidising microbes would not proliferate and the pseudo-first-order rate
constant would then be lower.
The simulation with an oxidation threshold based on this idea gives similar
\chem{[Mn_\text{diss}]} as the aggregation threshold; but as we saw in the previous section, it
would influence significantly the \chem{Mn_\text{ox}}/\chem{Mn_\text{diss}}
concentration ratio in the deep ocean, making for a less realistic simulation.
Therefore, this alternative should probably not be pursued.

Secondly, the process of oxidation is strongly mediated by adsorption
onto particles, after which they form larger aggregates that settle
faster.
This is now parameterised by increasing the settling velocity of
particles with depth.
However, a more explicit model for adsorption/desorption could be useful.
Possibly, much of the \chem{Mn_\text{ox}} does not settle efficiently.
In that case, another tracer of adsorbed Mn is needed that would export oxidised
Mn (in adsorbed or aggregated form) but not too fast to account for the
background \chem{[Mn_\text{diss}]}.
The adsorption should not take place onto a homogeneous pool of
particles as is effectively done in our model, but rather onto a
component, for example calcium carbonate (\chem{CaCO_3})
\citep{martin1983,book::silva2001} and lithogenic particles
\citep{roybarman2009}.
Sediment samples show a strong correlation between authigenic manganese
and lithogenic particles, though Mn does not show any correlation with
biogenic silica, POC or \chem{CaCO_3} according to recent observations
and modelling \citep{roybarman2009}.
On the one hand, this suggests that only lithogenic particles are a scavenger of
Mn.
On the other hand, we do not know whether the correlation comes from Mn
adsorption onto lithogenic particles, or if it is lithogenic in itself.

Thirdly, oxidation is mediated by microbes (mainly bacteria)
\citep{sunda1994,vonlangen1997,tebo2005,book::silva2001}.
Possibly they prolong the time of \chem{Mn_\text{ox}} particles spent in the euphotic zone, but
this has not been tested.
In the model we assume that these processes are included in the redox rate
constant.
If we want the redox rate constants to depend on inhomogeneously
distributed quantities like the bacterial distribution or the oxygen
concentration, the equations need to be modified, and
our model may need to be extended.
Subsequently, more Mn would stay suspended.

Fourthly, it has been suggested that dissolved Mn ligands keep Mn in
solution~\citep{sander2011,madison2013,luther2015}.
That would result in less conversion to manganese oxide aggregates, and
hence may solve the removal problem, but this would yield wrong \chem{Mn_\text{ox}}/\chem{Mn_\text{diss}}
concentration ratios like we saw for \textit{OxidThreshold}.
Similarly, nanoparticles, which have a size that falls within the operationally
defined dissolved Mn, keep Mn afloat as long as they do not aggregate.

\subsubsection{Hydrothermal activity}

In the model, hydrothermal \chem{Mn_\text{diss}} is added to the ocean
based on a model proxy of \chem{^3\!He} \citep{dutay2004}.
For the \textit{Reference} simulation the hydrothermal \chem{Mn_\text{diss}}
flux was set to \num{0.10e9} mole \chem{Mn} for each mole of \chem{^3\!He}.
There is no reference to literature values in Table~\ref{tab:params_Mn}
since these are very uncertain.
For instance,
\citet{klinkhammer1980} note that ``\citet{corliss1979} reported that
the \chem{Mn}\,:\,\chem{^3\!He} ratios vary by over a factor of 2 in the
Galapagos vents, and as we discuss below, similar variations are found
at 21\degree\,N\@''.
We have chosen a value to yield an acceptable distribution of \chem{Mn_\text{diss}}.

The goal of \textit{LowHydro} is to investigate whether the combination of the
high hydrothermal input of \chem{Mn_\text{diss}}, and strong aggregation is
needed to get a more accurate simulation of the distribution of
\chem{Mn_\text{diss}}, namely that of \textit{Reference}.
To this end, we chose to decrease hydrothermal input and also to
decrease the settling velocity in \textit{LowHydro}.
In \textit{LowHydro} the spatial distribution of \chem{[Mn_\text{diss}]}
becomes more homogeneous.
Neither the low background concentration nor the high hydrothermal vent
concentrations are reproduced any more (Fig.~\ref{fig:mang:LowHydro_transect}).
Decreasing hydrothermal input and the settling velocity by a smaller factor
than 10 results in a \chem{Mn_\text{diss}} distribution that is better than
\textit{LowHydro} but not as good as in \textit{Reference} (not presented).
In any case, the main result is that a high hydrothermal input very similar to
or higher than that used in the \textit{Reference} simulation is needed to
reproduce~\chem{[Mn_\text{diss}]}.

According to observations, a negligible amount of Mn from hydrothermal
vents reaches the surface ocean.
As \citet{bruland2006} put~it:
    ``The hydrothermal input of iron and manganese [...] is essentially all
    scavenged and removed in the deep sea prior to having a chance to mix back
    into the surface waters.''
However, this is what we have thought about iron while there are now
doubts about this \citep{tagliabue2010}.
Similarly, hydrothermal Mn may not be completely removed from the ocean before
reaching the surface by currents and vertical mixing.
There are different potential reasons for this.
The oxidation of Mn(II) is thermodynamically favoured, but the large
activation energy of Mn(II) oxidation renders Mn(II) stable in
aquatic environments \citep{nealson2006}.
Other potential reasons for \chem{Mn_\text{diss}} stability are already given on
the previous page.
The low \chem{Mn_\text{diss}} concentrations away from hydrothermal vents are
established in our model by removing \chem{Mn_\text{ox}} from the high
hydrothermal Mn input down to \SI{25}{\pico\molar}.

Interestingly, in this way, in \textit{Reference}, hydrothermal flux accounts
for 92\,\% of the total Mn input.
Dust deposition is 18 times and sediment flux 31 times as small as hydrothermal input, and still
\chem{[Mn_\text{diss}]} in the upper 1000\;m of the ocean is dominated by
\chem{Mn_\text{diss}} release from marginal sediments and dissolution from
deposited dust.
Nonetheless, some of the hydrothermally derived \chem{Mn_\text{diss}} reaches
the ocean surface.
In our model, 8\,\% of \chem{Mn} in the surface is hydrothermal of origin.
The reason for the importance of sediment and dust is that in the model the
settling velocity of \chem{Mn_\text{ox}} is much higher near hydrothermal vents
(up to \SI{10}{\metre\per\day}) than in the surface ocean
($w_s=\SI{1}{\metre\per\day}$).
The modelling study by \citet{lavelle1992} suggests a
settling velocity much higher than our $w_\text{ox}$.
They used a model with an additional tracer of large particles
(aggregate products) that had settling speeds of up to
\SI{175}{\metre\per\day} (near hydrothermal vents, as they model those
regions).
This shows that using an increasing settling velocity is justified.
Moreover, it suggests that at least two particle tracers are needed in
the model.
It is worth investigating in future studies if this would make for a more
accurate and reliable model.

\conclusions

This is the first study in which the 3-D distribution of \chem{Mn_\text{diss}}
has been modelled and compared to the recent observations from the
\textsc{Geotraces} programme.
A combination of photoreduction and Mn sources to the upper ocean yields
high surface concentrations of dissolved manganese.
However, the concentration is at several locations underestimated by the model.
The Mn sources to the ocean surface and the sinks are uncertain and could
therefore be adjusted for a more accurate simulation.
The most important sources for the upper ocean are sediments, dust, and,
more locally, rivers, whereas hydrothermal vents are the most important in the
deep ocean.
The observed sharp hydrothermal signals are produced by assuming both a
strong source and a strong removal of Mn near hydrothermal vents.
Our model further shows that the Mn at the surface in the Atlantic Ocean moves downwards
into the North Atlantic Deep Water, but because of
strong removal the Mn signal does not propagate southwards.

There is a mainly homogeneous background concentration of dissolved Mn
of about \SIrange{0.10}{0.15}{\nano\molar} throughout most of the deep ocean.
Our model reproduces this by means of a threshold on solid manganese oxides of
\SI{25}{\pico\molar},
suggesting that a minimal concentration of particulate Mn is needed before
aggregation and removal become efficient.
An aggregation threshold, as applied in our model, appears reasonable, and does
not affect the modelled \chem{Mn_\text{ox}}/\chem{Mn_\text{diss}} concentration ratio.
An oxidation threshold is more troublesome as it affects this ratio in an
undesirable way.
Since the settling condition as a sole threshold already appears to mitigate most
of the removal, it is reasonable to further develop the model with the
aggregation threshold and without the oxidation threshold.
Still, the simplified redox and subsequent settling of \chem{Mn_\text{ox}} is incomplete
and could be too imprecise, as future studies might~show.

Our model includes biology, and this has a big impact in the Pacific Ocean.
However, we have at the moment no clear evidence for typical
uptake-remineralisation processes as for iron.
At the same time, we are missing dominant sources from anoxic sediments,
especially in upwelling regions like at the East Pacific boundary.
Those sources, once adequately represented in the simulation, might be large
enough to compensate for the effect of our biological Mn uptake at the surface.
Such a source, and possibly
biological incorporation, should be improved in future versions of the model.
Besides these processes, it may also be needed to
model microbial activity throughout the ocean, as that is not
homogeneously distributed.
The oceanwide chemical first-order reaction may be inadequate to represent such
microbial activity in certain regions.

Hydrothermal fluxes of \chem{Mn} were set to such a high rate that we must
assume 96\,\% is scavenged near the outflow of the vents.
This choice was made to account for the local high \chem{Mn_\text{diss}} near the oceanic ridges.
This was combined with a high $w_\text{ox}$ in the deep ocean to prevent
hydrothermal \chem{Mn_\text{diss}} spreading far away from the oceanic ridges in
a too high concentration.
As an alternative, settling may not need to be set as high as
\SI{10}{\metre\per\day} in the deep ocean, but then it is unclear how to
simulate the local nature of \chem{[Mn_\text{diss}]} anomalies near the
ocean ridges.
One possibility is to include extra species of \chem{Mn}, which may include a very
fast sinking particle and a very reactive \chem{Mn} species.

Process studies of \chem{Mn} are necessary to determine the rate constants, and
possibly thresholds, for redox, scavenging and aggregation.
More measurements of particulate \chem{Mn_\text{ox}} concentrations would be useful as well.
When measuring particulate Mn, it needs to be clear what is measured exactly.
It would for instance be useful if (1) a range of particle sizes were
measured, and (2) a structural analysis of the particles were performed,
such that one can unambiguously say onto which particle \chem{Mn} were adsorbed
or into which particle it were incorporated.

\appendix

\setcounter{figure}{0}
\renewcommand{\thefigure}{A\arabic{figure}}
\setcounter{table}{0}
\renewcommand{\thetable}{A\arabic{table}}
{\small
\section{Data--model comparison}        \label{sec:comparison}
We compare quantitatively our model results with the observations at the GA02
\textsc{Geotraces} transect in the West Atlantic Ocean.
First the model output is horizontally interpolated onto the station
coordinates, keeping the vertical model grid (\SI{10}{\metre} at the surface up
to \SI{500}{\metre} thickness near the bottom).
Then the modelled \chem{[Mn_\text{diss}]} that lie closest to each of the
observations are associated with each other.
To be precise, for each observation, we take the shallowest gridbox whose upper
bound lies deeper than the observation,
after which residuals can be defined as $P_i-O_i$,
where $O_i$ is the observed and $P_i$ the modelled \chem{[Mn_\text{diss}]},
for each $i \in \{1, ..., N\}$, with $N$ the number of data points.
The interpolation introduces a representative error that is taken to be
part of the residual \citep[][]{thesis::vanhulten2014}.
Then several statistics are determined, namely the Root Mean Square
Deviation (RMSD), the Reliability Index (RI) and the Pearson correlation
coefficient~$r$.

%

In addition to classical statistical indices (Pearson correlation index
\textit{r}, root mean square error RMS), another
additional performance indicator has been used as suggested in previous skill
assessment studies \citep{stow2009,vichi2009}: the reliability index
\citep{leggett1981}.
The reliability index ``quantifies
the average factor by which model predictions differ from observations''
\citep{stow2009}.
It is in essence the root-mean-square deviation, but it uses the logarithm of
the residual.
This is useful when both large and small values need to be considered (as for
this case).
The RI is given by
\begin{equation}
    \mathit{RI} = \exp{{\sqrt{\frac{1}{N} \sum_{i=1}^N\left(\log\frac{O_i}{P_i}\right)^2}}} \,.
\end{equation}
where $O_i$ is the measured concentration with index $i$, $P_i$ is the model
prediction associated to the respective observation $i$, and
$N$ is the number of observations.

Furthermore, for each sensitivity simulation the significance of the
change in each goodness-of-fit statistic compared with the corresponding
reference simulation is calculated.
This is determined by means of a Monte Carlo simulation on the reference
simulation for which a subsample of 400 is randomly selected from
the original set of 1320 data--model points.
They are the pairs of observations and model output, both on the model
grid.
This is done 50\,000 times, and from this the $2 \sigma$ confidence
interval is calculated (the mean $\pm$ two times the standard
deviation).
Suppose that we want to simulate $q$, and assume $q$ is in steady state.
Given a reference simulation $X$, for any model simulation $Y$ resulting in
$q_Y(\mathbf{x})$, the RMSD of
$q_Y(\mathbf{x})$ must be outside the $2 \sigma$ confidence range of the
RMSD distribution of $q_X(\mathbf{x})$ to say that $Y$ is a significant
improvement or worsening compared to~$X$.

The statistics tell us how high the modelling accuracies of our simulations
are
Moreover, these statistics are illustrated with comparison plots and an
extensive discussion of those.
For the visual comparison between model and observations, horizontal and
vertical cross-sections of the model data are presented.
Using the same colour scale, observations are plotted as coloured dots
to directly compare the model with the observations.
Horizontal \chem{[Mn_\text{diss}]} sections are presented for four
different depths, where ``surface'' signifies the average over the upper
\SI{45}{\metre}, ``\SI{500}{\metre}'' is 400--600\;\si{\metre} averaged,
``\SI{2500}{\metre}'' is 2100--2900\;\si{\metre} averaged and
``\SI{4500}{\metre}'' is 4000--5000\;\si{\metre} averaged.
The colour scale is not linear to better show the main features at both
low and high concentrations of \chem{Mn_\text{diss}}.
The vertical \chem{[Mn_\text{diss}]} sections are calculated from the
3-D model data by converting the ORCA2 gridded model data to a
rectilinear mapping and interpolating the rectilinear data onto the
cruise track coordinates.

\section{Code availability}             \label{sec:code}

We used NEMO~3.6 svn~7036 that can be downloaded through this command:
\begin{verbatim}
$ svn checkout -r 7036 \
http://forge.ipsl.jussieu.fr/nemo/svn/branches/2015/nemo_v3_6_STABLE
\end{verbatim}
However, the latest stable version of NEMO is usually advisable.
A login will be asked that you can create on the
\href{http://www.nemo-ocean.eu/}{NEMO website}.
The manganese module comes as part of different Fortran~90/95 source code files
that are to be found in the electronic supplement.
Those should be put in \texttt{NEMOGCM/CONFIG/*/MY\_SRC/} that adds and
overrides symlinks in \texttt{NEMOGCM/CONFIG/*/WORK/}.
The asterisk must be replaced with your configuration name that should be based
on \texttt{ORCA2\_OFF\_PISCES}.

Both NEMO and the manganese model are available under the
\href{http://www.cecill.info/}{CeCILL licence}.

Model output data are available at \url{https://doi.org/10.1594/PANGAEA.871981}.
}

\begin{acknowledgements}
We would like to thank several people in particular who helped in different ways in
this study.
Angela Milne, William Landing and Joseph Resing kindly provided their manganese data from the
Pacific Ocean.
We thank Caroline Slomp, Catherine Jeandel and Micha Rijkenberg for the
useful discussion.
Loes Gerringa has kindly provided speciation calculations on manganese.
Loes Gerringa and Micha Rijkenberg organised and served as expedition leaders of
the three cruises of the GA02 section.
This research was funded by the \href{http://www.nwo.nl/en/}{NWO} (grant
839.08.410; \textsc{Geotraces}, Global Change and Microbial Oceanography in the
West Atlantic Ocean and grant 820.01.014 \textsc{Geotraces} Netherlands-USA
Joint Effort on Trace Metals in the Atlantic Ocean).
This study was partly supported by a Swedish Research Council grant
(349-2012-6287) in the framework of the French-Swedish cooperation in the common
research training programme in the climate, environment and energy agreement
between VR and LSCE, for the project ``Particle transport derived from isotope
tracers and its impact on ocean biogeochemistry: a \textsc{Geotraces} project in the
Arctic Ocean''.
The sampling and analysis of the data of \citet{data::milne:p16} were supported
by \href{http://www.nsf.gov/}{NSF} grants OCE-0223378, OCE-0550317 and
OCE-0649639.

\end{acknowledgements}

%
\DeclareRobustCommand{\DutchName}[4]{#1, #3~#4}

\bibliography{papers_oceano,theses_oceano,books_oceano,present_oceano,data_oceano}{}
\bibliographystyle{copernicus}
\end{document}